%% file: Arxiv.tex
\documentclass[lettersize,journal]{IEEEtran}

\IEEEoverridecommandlockouts
\usepackage[T1]{fontenc}
\newif\ifhavebib

\havebibtrue
\usepackage{latexsym}
\usepackage{setspace}
\usepackage{bm}
\usepackage{lipsum}
\usepackage{bbm}
\input{Chead.tex}

\input{defns.tex}

\begin{document}
\providecommand{\keywords}[1]{\textbf{\textit{Index terms---}} #1}
\title{Differentially Private Distribution Release of Gaussian Mixture Models via KL-Divergence Minimization}
\author{
		Hang Liu,~\IEEEmembership{Member,~IEEE}, Anna~Scaglione,~\IEEEmembership{Fellow,~IEEE}, and Sean Peisert,~\IEEEmembership{Senior Member,~IEEE}
		\thanks{
  This research was supported in part by the Director, Cybersecurity, Energy Security, and Emergency Response (CESER) office of the U.S. Department of Energy, via the Privacy-Preserving, Collective Cyberattack Defense of DERs project, under contract DE-AC02-05CH11231.  Any opinions, findings, conclusions, or recommendations expressed in this material are those of the authors and do not necessarily reflect those of the sponsors of this work. 
  
  Hang Liu is with the State Key Laboratory of Internet of Things for Smart City and the Department of Electronic and Communication Engineering, University of Macau, Macao S.A.R. (email: hangliu@um.edu.mo). Anna Scaglione is with the Department of Electrical and Computer Engineering, Cornell Tech, Cornell University, New York, NY, 10044 USA (e-mail: as337@cornell.edu). Sean Peisert is with the Computing Sciences Research, Lawrence Berkeley National Laboratory, Berkeley, CA 94720 USA (e-mail: sppeisert@lbl.gov).
		}}
\IEEEoverridecommandlockouts
\maketitle
\begin{abstract}
Gaussian Mixture Models (GMMs) are widely used statistical models for representing multi-modal data distributions, with numerous applications in data mining, pattern recognition, data simulation, and machine learning. However, recent research has shown that releasing GMM parameters poses significant privacy risks, potentially exposing sensitive information about the underlying data. In this paper, we address the challenge of releasing GMM parameters while ensuring differential privacy (DP) guarantees. Specifically, we focus on the privacy protection of mixture weights, component means, and covariance matrices. We propose to use Kullback-Leibler (KL) divergence as a utility metric to assess the accuracy of the released GMM, as it captures the joint impact of noise perturbation on all the model parameters. We develop a KL-divergence-driven optimization framework for differentially private GMM release that jointly designs privacy-budget allocation and parameter perturbations. Through theoretical analysis, we quantify the effects of privacy budget allocation and perturbation statistics on the DP guarantee, and derive a tractable expression for evaluating KL divergence. We formulate and solve an optimization problem to minimize the KL divergence between the released and fitted non-private models, subject to a given $(\epsilon, \delta)$-DP constraint. Extensive experiments on both synthetic and real-world datasets demonstrate that our approach achieves strong privacy guarantees while maintaining high utility.
\end{abstract}
\section{Introduction}
In recent years, the remarkable success of data-driven artificial intelligence (AI) has spurred an increasing demand for the sharing and analysis of large-scale, multi-class, and high-dimensional datasets across a variety of domains, such as healthcare records, consumer transactions, and mobility traces. Organizations have recognized the potential of sharing data statistics to enhance data mining, improve public services, optimize recommendations, and facilitate data simulation \cite{bishop2006pattern}. However, sharing raw data or even their statistics raises significant privacy concerns, especially when sensitive attributes of individuals might be inferred, underscoring the need for privacy-preserving mechanisms that allow the release of data statistics without exposing private information \cite{dwork2006calibrating}. 

In this paper we take the view that releasing a differentially private generative model enables arbitrary downstream analysis by allowing users to sample synthetic data, while preserving the same privacy guarantees for every derived study, optimization, or query.
Among the various statistical modeling approaches, we focus on the Gaussian Mixture Models (GMMs), which stand out as a versatile tool for representing complex, multivariate, and multi-modal data distributions \cite{figueiredo2002unsupervised}. GMMs are widely applied in data mining and machine learning (ML). By modeling the overall distribution as a mixture of several Gaussian components, GMMs naturally capture latent subgroups or clusters in the data, such as different risk profiles in healthcare or distinct spending behaviors in retail. Furthermore, GMMs provide a compact and interpretable parameterization, including component means, covariances, and mixture weights (also known as categorical frequencies), which can be shared more efficiently than raw data records. As a motivating real-world application, GMMs are accurate in representing the statistics of energy consumption data whose release is critical for grid modernization efforts; conveniently, the logarithmic values of load demand profiles in Advanced Metering Infrastructure (AMI) fit well to GMMs \cite{9799519}. In the context of energy usage, these load demand profiles represent the power consumption profile of a given user or household. 
{Sharing realistic generative models for real consumption data, that can be sampled and queried for arbitrary downstream tasks, opens the door for third parties and utilities vendors to perform accurate planning studies, policy makers to better regulate utilities and empowers advocates to potentially help customers better understand or dispute their energy bills}. 

{\bf Threat model}: As in any release of statistical aggregates of sensitive data, releasing GMM parameter estimates that fit a specific sample-set poses privacy risks, since repeated queries for such parameters leak information about the samples used to compute the estimates. An adversary could potentially exploit small mixture components or extreme values, to infer specific individuals’ records or their associated class labels. For instance, changes in the means of each of the GMM that come from asking to change the samples used in estimating the parameters can identify the presence of a specific customer. 

To mitigate these risks, differential privacy (DP) \cite{dwork2006calibrating,dwork2014algorithmic} offers a robust framework to quantify and control the privacy loss resulting from the inclusion or exclusion of any individual record in a dataset. By introducing carefully calibrated noise, we can estimate mixture model parameters that satisfy strong privacy guarantees while still enabling meaningful statistical analysis. 
A central challenge in applying DP mechanisms to the GMM parameters fitting as sample set is maintaining fidelity between the fitted non-private distribution and the released one under a DP constraint. Conventional DP mechanisms typically add artificial noise to perturb the model parameters in order to protect data privacy, at the expense of accuracy of the released GMM. 
Existing work \cite{7590445} proposed the release of differentially private GMMs by adding white Laplace noise to the mixture weights, means, and covariances individually, measuring the accuracy of differentially private GMMs using the parameters' mean-squared errors (MSEs). 

Rather than the MSEs of the individual parameters' release, we argue that distribution divergence, specifically \emph{Kullback–Leibler (KL) divergence}, is another comprehensive metric for quantifying the fidelity of GMMs. KL divergence is a well-established divergence measure in information theory that quantifies how much one probability distribution diverges from another \cite{cover1999elements}. When used to characterize the utility of distribution release, KL divergence measures the ``information'' lost when replacing the fitted non-private distribution with a privacy-preserving model. It is well-known that computing maximum likelihood estimation (MLE) is asymptotically equivalent to minimizing the KL divergence between the empirical and true distributions \cite{DL_Goodfellow}. Thus, publishing a data distribution with minimal KL divergence ensures that the released model is as accurate as possible in the maximum-likelihood sense. Moreover, when applied to data simulation and augmentation for ML, KL divergence is closely related to the generalization ability of ML models. Extensive research has established PAC-Bayes bounds for quantifying the generalization error of ML models \cite{MAL-112}, which depends critically on the KL divergence between the training and testing data distributions. In this context, KL divergence serves as a powerful metric for evaluating how well an ML model trained on a given data distribution (e.g., a released GMM under privacy constraints) generalizes to the true data, thus providing an important tool for assessing the performance of data synthesis and augmentation in ML training.

Recognizing the importance of the KL divergence as a figure of merit for the parameters' release, in this work, we study how to minimize the KL divergence for GMM parameters' release while meeting a desired privacy budget for the dataset used in fitting the model.

\subsection{Our Contributions}
Motivated by the above discussions, in this work 
{\color{black}we develop a KL-divergence-driven optimization framework for differentially private GMM release. Specifically, we study the fitting of a multi-class labeled dataset to a GMM whose parameters, including mixture weights, Gaussian means, and covariances, are then released under a given DP requirement. Building on Gaussian and Wishart perturbations and standard DP composition, the framework jointly optimizes privacy-budget allocation, perturbation statistics, and a randomized mapping of mixture weights to minimize the expected KL divergence from the fitted non-private GMM. To this end, we derive a tractable expression to evaluate the KL divergence between the released private GMM and the fitted non-private model. We then present a method to bound the achievable DP level by carefully controlling the noise injection. Through this, we optimize the noise distributions to minimize KL divergence subject to a desired DP constraint.} 
The main contributions of this paper are summarized as follows:
\begin{itemize}
    \item \textbf{DP-GMM Model Release Framework:} We propose a two-step approach for the differentially private release of GMMs by fitting and releasing GMM model parameters under any given $(\epsilon,\delta)$-DP constraint. Our method first fits the dataset into a GMM and then injects \emph{multivariate} Gaussian and Wishart noise into the Gaussian means and covariances, respectively. For the discrete-valued mixture weights, we introduce a random mapping mechanism that maps the estimated weights to other feasible mixture weights with a controllable mapping probability. We then formulate the DP mechanism design problem as optimizing the privacy budget and the noise distributions to balance the trade-off between DP and data utility, measured by the KL divergence between the released GMM and the fitted non-private model.
    
    \item \textbf{Privacy Analysis and Privacy-Utility Trade-Off Characterization:} We conduct an $(\epsilon, \delta)$-DP analysis, bounding the combined privacy loss from releasing both the continuous parameters (means and covariances) and the discrete parameters (mixture weights). In addition, we derive a tractable expression for the KL divergence between the privatized GMM and the fitted non-private distribution. This enables us to characterize the trade-off between privacy and the overall fidelity of the released model, providing a clear path to control this balance.

    \item \textbf{DP Mechanism Design via Privacy-Constrained KL-divergence Minimization:} Based on our analysis of model utility and privacy, we formulate an optimization problem that allocates privacy budgets and optimizes noise distributions by minimizing the expected KL divergence subject to a given $(\epsilon, \delta)$-DP constraint. {Crucially, this formulation makes the privacy–utility trade-off \emph{transparent}: the privacy budget is reallocated across DP randomization parameters during the alternating updates, directly exposing how noise levels and budget shares affect the utility.} While the problem is non-convex, we propose a low-complexity alternating optimization solution to obtain a solution efficiently through alternating optimization. 
\end{itemize}

Through extensive experiments on both synthetic and real-world datasets, we demonstrate how our method preserves model accuracy while satisfying rigorous privacy criteria. Specifically, we show that our approach provides a differentially private GMM that adheres to stringent DP requirements, while maintaining a much lower KL divergence and preserving model fidelity in comparison to existing DP mechanisms.

\subsection{Organization and Notations}
The remainder of this paper is organized as follows. We introduce the system model for differentially private parameter release of GMMs in Section \ref{sec2}. In Section \ref{sec3}, we analyze the achievable DP and formulate the DP mechanism design problem.  In Section \ref{sec4}, we introduce the proposed solution to the DP mechanism design problem. In Section \ref{sec5}, we present experimental results to evaluate the proposed method. Finally, this paper concludes in Section \ref{sec6}.
	
Throughout, we use regular, bold lowercase, and bold capital letters to denote scalars, vectors, and matrices, respectively. We use  $\Xv^T$ to denote the transpose of $\Xv$, $\Xv^H$ to denote the conjugate transpose, $\tr(\Xv)$ to denote the trace, and $|\Xv|$ to denote the determinant of $\Xv$.
We use $x_i$ to denote the $i$-th entry of vector $\xv$, $x_{ij}$ or $X_{i,j}$ interchangeably to denote the $(i,j)$-th entry of matrix $\Xv$, and $\xv_j$ to denote the $j$-th column of $\Xv$. 
The real normal distribution with mean $\muv$ and covariance $\Cv$ is denoted by $\Norm(\muv,\Cv)$, and the cardinality of set $\mathcal{S}$ is denoted by $\abs{\mathcal{S}}$. We use $\norm{\cdot}_p$ to denote the $\ell_p$ norm, $\Iv_N$ to denote the $N\times N$ identity matrix,  $\bf 1$ (or $\bf 0$) to denote the all-ones (or all-zeros) vector with an appropriate size. We use $x\sim p(x)$ to represent that the random variable $x$ is drawn from the distribution $p(x)$.

\section{Differentially Private Distribution Release for GMMs}\label{sec2}
Consider a dataset $\mathcal{D} = \{(\xv_n, y_n)\}_{n=1}^N$, consisting of $N$ labeled samples $\xv_1, \dots, \xv_N \in \mathbb{R}^d$ partitioned into $K$ classes. Each sample $\xv_n$ is associated with a class label $y_n \in \{1, \dots, K\}$.\footnote{When unavailable, the class labels can be calculated \emph{a priori} by using clustering methods such as $K$-Means clustering.} We assume that every class in the considered dataset and in each admissible neighboring dataset contains more than two samples, i.e., $N_k>2$ for all $k$. A neighboring pair is admissible only when this condition holds at both endpoints. Our aim is to publicly release a distributional model for these multi-class data in a differentially private manner by fitting them with a \emph{multivariate} GMM.

Let $\pi_k(\mathcal{D})>0$ denote the empirical frequency (mixture weight) of class $k$ satisfying $\sum_k \pi_k(\mathcal{D})=1$, and $\muv_k(\mathcal{D})$ and $\Sigmav_k(\mathcal{D})$ represent the mean and covariance of the $k$-th Gaussian component, respectively; the GMM fitting $\mathcal{D}$ is such that, for each cluster $k$:
\begin{align}\label{eq01}
  p(y=k)=\pi_k(\mathcal{D}),  p(\xv|y=k) = \Norm\left(\muv_k(\mathcal{D}), \Sigmav_k(\mathcal{D})\right) .
\end{align}

To satisfy a prescribed DP constraint, we employ a \emph{two-step} strategy, by first estimating the parameters $\{\pi_k(\mathcal{D}), \muv_k(\mathcal{D}), \Sigmav_k(\mathcal{D})\}_{k=1}^K$ from the dataset $\mathcal{D}$, followed by applying a DP mechanism before releasing these parameters. Concretely, we compute the parameters via histograms and sample statistics as follows:
\begin{subequations}\label{eq2}
\begin{align}
&\pi_k(\mathcal{D}) = \frac{1}{N}\sum_{y_n=k} 1, \label{eq2a}\\
& \muv_k(\mathcal{D}) = \frac{1}{N\pi_k}\sum_{y_n=k} \xv_n, \\
& \Sigmav_k(\mathcal{D}) = \frac{1}{N\pi_k - 1}\sum_{y_n=k} (\xv_n - \muv_k)(\xv_n - \muv_k)^T+\kappa\Iv_d.
\end{align}
where a some constant $\kappa$ makes the covariance matrix positive definite.

\end{subequations}
Define $\piv = [\pi_1,\dots,\pi_K]$. We use the following set $\mathcal S$ as the output space for the released mixture weights; it also contains all input histograms satisfying $N_k>2$:
\begin{align}
    \mathcal{S}=\left\{\piv:\sum_{k=1}^K \pi_k=1,\pi_k\in\left\{\frac{1}{N},\frac{2}{N},\cdots,\frac{N-1}{N}\right\},\forall k\right\}.
\end{align}
By the stars-and-bars theorem, the cardinality of $\mathcal{S}$ is
given by $|\mathcal{S}|= {{N-1}\choose{K-1}}$. Unless otherwise specified, when we refer to an element $\piv\in\mathcal{S}$, we assume a fixed ordering of the elements in $\mathcal{S}$ and thus any feasible  $\piv$ has a unique index in the order.

\subsection{Threat Model and DP Definition} 
After estimating $\{\pi_k(\mathcal{D}), \muv_k(\mathcal{D}), \Sigmav_k(\mathcal{D})\}_{k=1}^K$, we apply a randomized sanitization mechanism to satisfy given DP constraints before making these parameters public. 

{\bf Model-release privacy threat}: a trusted curator fits a GMM to the raw dataset $\mathcal{D}$ and releases a privatized version of the model parameters to an untrusted recipient. A powerful adversary is assumed to observe the released parameters $(\widetilde{\piv}(\mathcal{D}),\{\widetilde{\muv}_k(\mathcal{D}),\widetilde{\Sigmav}_k(\mathcal{D})\}_{k=1}^K)$ and aims to infer whether the class label of a particular individual has a specific value (or changes between two candidate labels). Under our default \emph{label-level} definition, neighboring datasets $\mathcal{D}'$ differ by flipping exactly one label while holding the feature vectors fixed; thus the protected unit is a single label. This setting is motivated by applications where labels encode sensitive membership (e.g., consumer type, risk category, cluster assignment) while the trusted curator retains both features and labels locally and releases only privatized GMM parameters. Formally, the output distributions on adjacent datasets must satisfy the following standard definition of DP, introduced in \cite{dwork2006calibrating,dwork2014algorithmic}.

\definition[$(\epsilon,\delta)$-DP]{\label{def1}
Consider a randomized mechanism $M$ that takes a dataset $\mathcal{D}$ as input and outputs a query answer $M(\mathcal{D})$. Let $\epsilon>0$ and $\delta \in [0,1]$ be given parameters. $M$ is said to satisfy $(\epsilon,\delta)$-DP if, for any two admissible adjacent datasets $\mathcal{D}$ and $\mathcal{D}^\prime$ where $\mathcal{D}^\prime$ differs from $\mathcal{D}$ by altering the class label of exactly one data point, the following inequality holds for all measurable sets $\mathcal{H}$:
\begin{align}\label{DP_def}
   \Pr(M(\mathcal{D})\in\mathcal{H})\leq e^{\epsilon}\Pr(M(\mathcal{D}^\prime)\in\mathcal{H})+\delta, \forall \mathcal{H}.
\end{align}
}

In this work, $M$ refers to the mechanism that releases the GMM parameters $(\tilde{\piv}(\mathcal{D}),\{\tilde{\muv}_k(\mathcal{D}),\tilde{\Sigmav}_k((\mathcal{D}))\}_{k=1}^K)$. Intuitively, an $(\epsilon,\delta)$-DP mechanism makes it difficult for any potential attacker to detect the class label of any single data point in $\mathcal{D}$, especially when $\epsilon$ and $\delta$ are small. Here, $\epsilon$ controls the overall privacy stringency, while $\delta$ governs the probability that this privacy guarantee may not hold. {\color{black}We note that holding feature vectors fixed in Definition~1 specifies the adjacency relation and does not assume that the raw features are publicly available. The trusted data owner retains both features and labels locally and releases only the privatized GMM parameters. The label-level guarantee controls the information about individual labels disclosed by this model release, including when an adversary has prior knowledge of feature values. Keeping raw features local specifies the release protocol; protecting feature values through DP requires an adjacency relation that permits feature changes.}

\remark{{\color{black}Definition \ref{def1} adopts label-level DP, where an adjacent dataset $\mathcal{D}^\prime$ alters exactly one class label. Record-level DP is also possible under bounded features, with adjacency defined by adding, removing, or replacing one data point. Section \ref{sec4c} shows how our approach adapts to record-level DP with the same design principles. }
Note that under our default \emph{label-level} definition, neighboring datasets differ by flipping exactly one label while holding the feature vectors fixed; thus the protected unit is a single label. This setting is motivated by applications where labels encode sensitive membership (e.g., consumer type, risk category, cluster assignment) while the trusted curator retains both features and labels locally and releases only privatized GMM parameters. We show in Section \ref{sec4c} that the \emph{label-level} DP is technically the most demanding in our setting; once it is handled, adapting the method to record-level adjacency requires only minor and principled modifications.
}

\subsection{DP Approach}
{For brevity, since the dataset $ \mathcal{D} $ is implied by context in many of the following derivations we omit it as an argument in the notation.}
To satisfy a given DP requirement, we leverage well-known DP mechanisms that add artificial noise to $\{\pi_k,\muv_k,\Sigmav_k\}_{k=1}^K$ prior to release, with the caveat that we must ensure that the added noise respects the dimensionality and retains the feasibility of these parameters. Concretely, we need the perturbed frequency parameters to remain in $\mathcal{S}$, and the perturbed covariance matrices to be positive semidefinite (PSD).
Toward this end, we assign each parameter a specific noise distribution. For the continuous-valued mean and covariance of each Gaussian component, we adopt a \emph{multivariate} Gaussian mechanism and a Wishart mechanism \cite{Jiang_Xie_Zhang_2016}. The perturbed means and covariances are given by
\begin{align}  
    &\widetilde{\muv}_k= \muv_k+\wv_k,\wv_k\sim\Norm({\bf 0},\Gammav_k^{-1}),\forall k,\label{eq5}\\
    &\widetilde\Sigmav_k= \Sigmav_k+\Wv_k, \Wv_k\sim \mathcal{W}_d(\frac{1}{\gamma_k}\Iv_d,d+1),\forall k,\label{eq6}
\end{align}
where $\wv_k$ and $\Wv_k$ are the noise terms added to the mean and covariance of the $k$-th component, respectively, $\Gammav_k$ is the precision matrix of the Gaussian noise, $\mathcal{W}_d(\gamma_k^{-1}\Iv_d,d+1)$ is a Wishart distribution over $d\times d$ matrices with $d+1$ degrees of freedom and scale matrix $\gamma_k^{-1}\Iv_d$, and the parameter $\gamma_k>0$ controls the variance of the Wishart noise. As Wishart samples are always positive definite matrices, the construction in \eqref{eq6} ensures that the perturbed covariance remains feasible. Meanwhile, to privatize the discrete-valued frequency vector $\piv$, we release a random vector $\widetilde\piv \in \mathcal{S}$ according to a tunable conditional probability mass function (PMF) $f(\widetilde\piv | \piv)$. With a fixed ordering of elements in $\mathcal{S}$, this PMF can be viewed as a transition matrix $\Fv \in [0,1]^{|\mathcal{S}|\times |\mathcal{S}|}$ whose entry $F_{i,j}$ specifies the probability of mapping the $j$-th input element in $\mathcal{S}$ to the $i$-th output element in $\mathcal{S}$. In the sequel, we will use $f(\widetilde\piv | \piv)$ and $\Fv$ interchangeably to denote the mapping PMF. 

\remark{
Our randomization mechanism for the frequency parameter $\widetilde\piv$ presumes a discrete-valued input $\piv$, matching the histogram-based parameter estimation in \eqref{eq2a}. Other GMM estimation methods, such as the expectation-maximization algorithm \cite{dempster1977maximum} may yield continuous-valued frequencies, which can be discretized beforehand if needed.
{Note that the discretization resolution reflects the trade-off between accuracy and computational complexity. In our setting, $\mathcal{S}$ is the natural $1/N$-resolution histogram grid implied by Eq.~(2a), with $|\mathcal{S}|=\binom{N-1}{K-1}$. A finer grid (larger $N$) reduces discretization error and can improve utility, but it also increases $|\mathcal{S}|$ and the complexity of computing a full transition matrix.}
}

Our goal is to choose $\{\Gammav_k,\gamma_k\}_{k=1}^K$ and $\Fv$ so that the released parameters $\{\tilde\pi_k,\widetilde\muv_k,\widetilde\Sigmav_k\}_{k=1}^K$ satisfy the $(\epsilon,\delta)$-DP. An end user can then reconstruct the private GMM as
\begin{align}\label{eq07}
    \widetilde p(y=k)=\tilde\pi_k,\widetilde p(\xv|y=k)=\Norm(\widetilde\muv_k,\widetilde\Sigmav_k), \forall k.
\end{align}
We measure release fidelity using the expected KL divergence between the joint feature--label distributions:
\begin{align}\label{eq08}
    \E\left[\mathrm{KL}\left( \widetilde p(\xv,y)\| p(\xv,y)\right)\right]=\E\left[\sum_{k=1}^K\int \widetilde p(\xv,k) \ln \Big(\frac{\widetilde p(\xv,k)}{p(\xv,k)}\Big) d\xv\right],
\end{align}
where the expectations are taken with respect to the randomness in $\{\wv_k,\Wv_k\}_{\forall k}$ and $\widetilde \piv$. A smaller KL divergence indicates that the released GMM better preserves the fitted non-private distribution. We note that {\color{black}the reference distribution $p(\xv,y)$ is the non-private GMM fitted to the observed dataset. Thus, the expected KL divergence above measures the additional distortion caused by privatization relative to this fitted model. Fidelity to the underlying data-generating distribution also depends on model selection and estimation accuracy; minimizing \eqref{eq08} preserves the fitted model but does not remove its approximation or estimation error.} 

The following examples demonstrate the effectiveness of the KL divergence for measuring the accuracy of distribution estimation and release.
\example[Classification]{
Classification is a fundamental problem in data mining and ML. By fitting a GMM to the raw data, one can capture latent structures or subgroups and use the model to classify new data. However, if the private GMM diverges significantly from the fitted non-private distribution, these classifiers may be distorted or merged incorrectly. For instance, consider a hospital that privately releases classification information about patients based on clinical measurements (high/low blood pressure, high/low cholesterol, etc.). Policymakers rely on accurate classification boundaries to identify high-risk subpopulations and allocate resources effectively. A high KL divergence between the fitted non-private and released distributions would imply that the separations are significantly altered, undermining the utility of the published model. Hence, minimizing the KL divergence in \eqref{eq08} ensures that the core data structure is well preserved despite the noise added for privacy.
}

\example[Density Estimation and Data Simulations]{
GMMs are widely used to estimate data density and generate synthetic samples that mimic the statistical properties of the real data. KL divergence is a standard metric for accuracy assessment and model selection of density estimation. Consider a public health agency creating a synthetic version of a patient dataset for open research. If the released GMM has a distribution that significantly diverges from the fitted non-private model in terms of KL divergence, the synthetic data may incorrectly represent disease prevalence or demographic proportions. This leads to flawed insights, misleading model development, and potential misallocation of medical resources. By monitoring and minimizing the KL divergence, the agency preserves the distributional patterns captured by the fitted non-private GMM, while still respecting strict privacy requirements.
}

\example[Data Augmentation for ML]{
Data augmentation plays a critical role in preventing model overfitting and enhancing model generalization, especially in applications where labeled data are expensive to obtain. A differentially private GMM allows organizations to share model parameters that can be used for sampling additional training data while preserving individual privacy. In machine learning theory, PAC-Bayes methods are commonly used to quantify generalization ability when training and testing data follow different distributions. Typical PAC-Bayes bounds indicate that the model generalization error is bounded by a non-diminishing term related to the KL divergence between the training and testing data distributions \cite{MAL-112}.

Moreover, in the context of data augmentation or synthesis, \cite{NEURIPS2023_a94a8800} reported that the generalization error of a model trained with an augmented or synthetic dataset is upper-bounded by a term determined by the total variation distance (TVD) between the true data distribution and the augmented data distribution. However, this TVD-based bound is often intractable, even for simple distribution models such as GMMs. According to Pinsker’s inequality \cite{csiszar2011information}, the TVD between two distributions is tightly upper-bounded by the square root of half their KL divergence. In other words, KL divergence serves as a surrogate for TVD and provides a more tractable expression for bounding the model generalization ability. In our setting, minimizing the KL divergence preserves the fitted non-private distribution used for data synthesis; the downstream classification experiments assess the resulting predictive performance.
}

{ \textbf{Scope (model-release privacy).} Our goal is to protect \emph{model-release} privacy rather than \emph{sample-level prediction} privacy. We assume a trusted data owner fits the GMM on the raw dataset and then releases differentially private parameters $\{\muv_k,\Sigmav_k,\piv\}$ to an untrusted query recipient. The DP guarantees we provide apply to the act of releasing these parameters.  Addressing per-query or per-sample prediction privacy is orthogonal to this work and left for future study.}
\section{Problem Formulation}\label{sec3}
In this section, we present a tractable problem formulation for DP mechanism design by minimizing the KL divergence subject to DP constraints.

\subsection{DP Analysis}
We begin by examining the conditions under which releasing $\{\tilde\pi_k,\widetilde\muv_k,\widetilde\Sigmav_k\}_{k=1}^K$ achieves a given $(\epsilon,\delta)$-DP requirement, as a function of the controllable parameters $\{\Gammav_k,\gamma_k\}_{\forall k}$ and $f(\widetilde\piv\mid \piv)$. The core idea is to bound the overall privacy loss by composing the individual privacy contributions of each parameter. Following the analysis in \cite{9799519,Jiang_Xie_Zhang_2016}, we derive a sufficient condition for satisfying \eqref{DP_def} when releasing $\{\widetilde\muv_k,\widetilde\Sigmav_k\}_{k=1}^K$. We then enumerate all possible adjacent datasets to derive the condition in \eqref{DP_def} for $\widetilde\piv$.

Given the dataset $\mathcal D$ of size $N$, let $\mathcal D^\prime_{n,k^\prime}$ denote the dataset obtained by changing the original label to $k^\prime$ while keeping all feature vectors fixed. 
There are at most $N(K-1)$ admissible adjacent datasets. A sufficient condition for satisfying $(\epsilon,\delta)$-DP is summarized below.

\theorem{\label{theorem1}
Consider the non-private parameters $\{\pi_k, \muv_k, \Sigmav_k\}_{k=1}^K$ defined in \eqref{eq2}. Releasing the perturbed parameters $\{\tilde\pi_k,\widetilde\muv_k,\widetilde\Sigmav_k\}_{k=1}^K$ in \eqref{eq5} and \eqref{eq6} satisfies $(\epsilon,\delta)$-DP if $\epsilon$ and $\delta$ fulfill the following inequalities:
\begin{align}
&\epsilon\geq 2\max_{k=1}^K\left\{\epsilon_k+\epsilon_k^\prime\right\}+\epsilon_0,\label{eq09}\\
    &\frac{\epsilon_k^2}{2\ln (4/\delta)}\nonumber\\
    &~~\geq \sup_{n,k^\prime\neq k} \left\{(\muv_k(\mathcal{D})- \muv_k(\mathcal{D}^\prime_{n,k^\prime}))^T\Gammav_k( \muv_k(\mathcal{D})- \muv_k(\mathcal{D}^\prime_{n,k^\prime}))\right\},\label{eq10}\\
   &\epsilon_k^\prime\geq \frac{3\gamma_k}{2N_k},\forall k,\\
    &\epsilon_0\geq \left|\ln\frac{f(\widetilde\piv|\piv(\mathcal{D}))}{f(\widetilde\piv|\piv(\mathcal{D}^\prime_{n,k^\prime}))}\right|,\forall \widetilde \piv\in\mathcal{S}, \forall n, k^\prime,\label{eq11}
\end{align}
where $N_k$ is the number of data points in class $k$, $\epsilon_0\geq0$ and $\epsilon_k\geq 0,\,1\leq k\leq K,$ are auxiliary variables. Note that in \eqref{eq11}, {the likelihood-ratio bound is quantified over \emph{all} possible outputs $\tilde{\piv}\in\mathcal{S}$, i.e., the full histogram support, rather than only over a dataset-dependent subset.} 

\begin{IEEEproof}
    See Appendix \ref{appa}.
\end{IEEEproof}
}

{
Intuitively, Theorem~\ref{theorem1} indicates that the global privacy guarantee (in terms of $\epsilon$) can be controlled by combining the privacy costs of each parameter release. The privacy-loss terms $\{\epsilon_k\}_{k=1}^K$, $\left\{\epsilon_k^\prime\right\}_{k=1}^K$, and $\epsilon_0$ characterize the sensitivity/DP cost of releasing each sample mean, sample covariance, and the mixture weights, respectively. {\color{black}Under a label flip, the source and destination class releases compose sequentially, giving the bound $2\max_k\{\epsilon_k+\epsilon_k^\prime\}$; composing with the mixture-weight release adds $\epsilon_0$. Each affected Gaussian mean uses failure probability $\delta/2$, giving total failure probability $\delta$.}
From \eqref{eq10}--\eqref{eq11}, we see that these sensitivity bounds are obtained by bounding the worst-case event, whose forms depend on the realized data values (or clipped values, when applicable), rather than on any assumption on the data distribution. We also note that the $(\epsilon,\delta)$ bound in Theorem~\ref{theorem1} depends explicitly on the hyperparameter $K$ and on the perturbation/noise variables $\{\Gammav_k,\gamma_k\}_{k=1}^K$ and $f(\tilde{\piv}|\piv)$. This dependence motivates jointly optimizing the privacy-allocation variables together with these perturbation variables.
}
This result gives us an explicit way to allocate the privacy budget $\epsilon_0,\epsilon_k,\forall k$ across the different noise additions, and then optimize the noise statistics $\{\Gammav_k,\gamma_k\}_{\forall k}$ and $f(\widetilde\piv| \piv)$ accordingly.\footnote{{Theorem~\ref{theorem1} relies on basic DP compositions, which enable a simple privacy-budget constraint and support a low-complexity solution. More advanced accountants, e.g., Rényi DP \cite{mironov2017renyi}, may tighten privacy bounds but lead to a different formulation and algorithmic structure. A full RDP-based development is an interesting direction for future work.
}}

\subsection{DP-Constrained KL-divergence Minimization}
We now formulate an optimization problem for the DP mechanism design. Our goal is to allocate the privacy budget $\epsilon_0$ and $\epsilon_k,\forall k$, and subsequently choose the noise statistics $\{\Gammav_k,\gamma_k\}_{\forall k}$ and $f(\widetilde\piv | \piv)$ so as to minimize the expected KL divergence in \eqref{eq08}, subject to a given $(\epsilon,\delta)$-DP constraint. The following proposition provides a tractable representation of the objective function in terms of these design variables.

\proposition{\label{theorem2}
The expected KL divergence in \eqref{eq08} can be rewritten as \eqref{eq_obj}, shown at the top of the next page. {The remaining log-determinant expectation is taken over the Wishart noise in \eqref{eq6} and can be evaluated numerically.}
}

\begin{IEEEproof}
We first decouple the objective into a mixture-weight loss and a within-component Gaussian KL-divergence term. The mean-noise quadratic term and the Wishart trace term admit closed-form expectations. The detailed derivation is provided in Appendix~\ref{appb}.
\end{IEEEproof}
\begin{figure*}
\begin{align}
   \eqref{eq08}&={\sum_{k=1}^K \E\!\left[\widetilde{\piv}_k \ln\frac{\widetilde{\piv}_k}{\piv_k}\right]
	+ \sum_{k=1}^K \E\!\left[\widetilde{\piv}_k \,\mathrm{KL}\!\left(\mathcal{N}(\widetilde{\muv}_k,\widetilde{\Sigmav}_k)\,\big\|\,\mathcal{N}(\muv_k,\Sigmav_k)\right)\right]}\nonumber\\
   &	{\color{black}=\sum_{\widetilde\piv\in\mathcal{S}} {f(\widetilde\piv|\piv(\mathcal{D}))}\underbrace{\sum_{k=1}^K \tilde\pi_k\left(\ln \frac{\tilde\pi_k}{\pi_k}+\frac{1}{2}\left(\frac{d+1}{{\gamma_k}}\tr(\Sigmav_k^{-1})+\tr(\Sigmav_k^{-1}{\Gammav_k}^{-1})-\mathbb E_{\Wv_k}\log\left|\Iv+\Sigmav_k^{-1/2}\Wv_k\Sigmav_k^{-1/2}\right|\right)\right)}_{\triangleq g(\widetilde\piv,\{\gamma_k,\Gammav_k^{-1}\}_{\forall k})}.\label{eq_obj}}
\end{align}
\hrulefill
\end{figure*}
{\color{black}\noindent\textbf{One-dimensional toy example.}
	To illustrate how the KL-divergence utility varies with the mean- and covariance-noise parameters, consider the simplest setting with scalar features ($d=1$) and one Gaussian component ($K=1$), where $p(x)=\mathcal N(\mu,\sigma^2)$, $w\sim\mathcal N(0,\Gamma^{-1})$, and $W=\chi_2^2/\gamma$. Here, $\chi_2^2$ denotes a chi-squared random variable with two degrees of freedom.For any noise realization $(w,W)$, the KL-divergence has a closed-form expression as
	$\mathrm{KL}(\mathcal N(\mu+w,\sigma^2+W)\|\mathcal N(\mu,\sigma^2))
	=\frac12\big(w^2/\sigma^2+W/\sigma^2-\ln(1+W/\sigma^2)\big)$.
	Taking expectations on both sides gives
	\begin{align}
		&\E[\mathrm{KL}(\mathcal N(\mu+w,\sigma^2+W)\|\mathcal N(\mu,\sigma^2))]\nonumber\\
		&=\frac12\left[
		\frac{\Gamma^{-1}}{\sigma^2}
		+\frac{2}{\gamma\sigma^2}
		-\E\ln\left(1+\frac{\chi_2^2}{\gamma\sigma^2}\right)
		\right],
	\end{align}
	The mean-noise contribution increases linearly with $\Gamma^{-1}$. The covariance contribution is strictly decreasing in $\gamma$ because the function $\gamma'-\ln(1+\gamma')$ increases for $\gamma'>0$, whereas $\chi_2^2/(\gamma\sigma^2)$ decreases with $\gamma$. It converges to zero as $\gamma\to\infty$, as required when the Wishart perturbation vanishes. This observation aligns with the intuition that with diminishing Gaussian and Wishart noise, the perturbed distribution converges to the non-private one. Thus, for fixed privacy-allocation variables, the best feasible $\gamma$ is its privacy-imposed upper bound; there is no unconstrained finite minimizer. The same qualitative behavior holds for multivariate GMM components: the mean-noise contribution remains linear in $\Gammav_k^{-1}$, while the covariance-noise contribution decreases with $\gamma_k$ and vanishes as $\gamma_k\to\infty$, consistent with the perturbed component approaching its non-private counterpart as both noise terms diminish.}

By Proposition \ref{theorem2}, we propose to solve the following KL-divergence minimization problem:
\begin{subequations}
\begin{align}
  (\text{P1}):  &\min_{\substack{\{\gamma_k,\Gammav_k,\epsilon_k\}_{k=1}^K\\\Fv,\epsilon_0}}   \eqref{eq_obj}\\
    &~~\text{s.t. } 2\max_{k} \left\{\epsilon_k+\frac{3\gamma_k}{2N_k}\right\}+\epsilon_0\leq \epsilon,\label{eq12b}\\
    &~~~~~~\eqref{eq10}, \eqref{eq11} \text{ hold},\label{eq12c} \\
    &~~~~~~\Gammav_k\succeq {\bf 0},\gamma_k>0,\epsilon_k>0,\forall k,\label{eq12d}\\
    &~~~~~~\epsilon_0\geq 0, \Fv^T{\bf 1}={\bf 1},[\Fv]_{i,j}\geq 0, \forall 1\leq i,j\leq |\mathcal{S}|.\label{eq12e}
\end{align}
\end{subequations}
Constraints \eqref{eq12b} and \eqref{eq12c} ensure the $(\epsilon,\delta)$-DP requirement, while \eqref{eq12d} and \eqref{eq12e} enforce feasibility of the precision matrices and the transition PMF. Note that (P1) is a non-convex problem and involves $\mathcal{O}(|\mathcal{S}|^2)$ constraints, posing a challenge for large $|\mathcal{S}|$.

        { In (P1), we optimize the privacy budgets $\{\epsilon_k\}_{k=1}^K$ and $\epsilon_0$ \emph{together with} the mechanism parameters. The feasibility constraints explicitly depend on problem hyperparameters such as $K$, $d$, and class sizes $\{N_k\}$. Consequently, smaller $N_k$ tightens the DP constraints and the optimizer responds by adjusting $\epsilon_k$ and increasing the Gaussian noise covariance $\Gammav_k^{-1}$ and the Wishart scale $\gamma_k^{-1}$; larger $N_k$ has the opposite effect. This joint treatment induces an \emph{automatic}, data-aware allocation across classes and parameters, avoiding manual tuning and typically yielding a tighter KL divergence utility. We refer the reader to the numerical results in Section~\ref{sec5} for a detailed investigation of the impact of key system hyperparameters.}
        
\section{DP Mechanism Optimization}\label{sec4}
In this section, we present several modifications to (P1) that simplify the DP mechanism optimization. We then propose a low-complexity algorithm to compute a sub-optimal solution based on alternating optimization.

\subsection{Problem Simplification}
Recall that $|\mathcal{S}|= {{N-1}\choose{K-1}}$. The matrix $\Fv$ contains $\binom{N-1}{K-1}^2$ entries, which grows as $\Theta(N^{2K-2})$ for fixed $K$, making (P1) computationally prohibitive for large $N$. However, we note that while $\mathcal{S}$ may be large, the adjacent datasets $\{\mathcal{D}^\prime_{n,k^\prime}\}$ of $\mathcal{D}$ correspond to only a small number of possible frequency vectors $\{\piv(\mathcal{D}^\prime_{n,k^\prime})\}$. Specifically, altering the label of exactly one data point changes exactly two entries in $\piv(\mathcal{D})$. Motivated by this, we define the possible frequency vectors generated by all the adjacent datasets of $\mathcal{D}$ as
\begin{align}
 &\{\piv(\mathcal{D}^\prime_{n,k^\prime})\}=\mathcal{S}^\prime(\mathcal{D})\nonumber\\
 =&\big\{\piv(\mathcal{D}^\prime)\in\mathcal{S}:\norm{\piv(\mathcal{D}^\prime)-\piv(\mathcal{D})}_0=2,\nonumber\\
 &\quad\norm{\piv(\mathcal{D}^\prime)-\piv(\mathcal{D})}_1=\frac{2}{N}, N\pi_k(\mathcal{D}^\prime)>2,\ \forall k\big\}.\label{eq14}
\end{align}
It follows that $ \mathcal{S}^\prime(\mathcal{D})\subset\mathcal{S}$ and $ |\mathcal{S}^\prime(\mathcal{D})|\leq K(K-1)$. Since $|\mathcal{S}^\prime(\mathcal{D})|\ll |\mathcal{S}|$, 
{to simplify the problem in (P1) we parametrize the mapping PMF only over $\piv(\mathcal{D})$ and its neighbors $\mathcal{S}^\prime(\mathcal{D})$. However, directly zeroing the probability outside $\{\piv(\mathcal{D})\}\cup\mathcal{S}^\prime(\mathcal{D})$ yields dataset-dependent support and hence violates DP, since adjacent inputs would induce different supports. We therefore apply the following smoothing:
\begin{align}
f(\widetilde\piv \mid \piv)=
(1-\lambda)f^\prime(\widetilde\piv \mid \piv)+\lambda/|\mathcal{S}|,
\qquad \forall\widetilde\piv \in \mathcal{S},
\end{align}
with a sufficiently large $\lambda\in(0,1)$ denoting a fixed smoothing parameter, $|\mathcal{S}|={{N-1}\choose{K-1}}$, and $f^\prime$ supported on $\mathcal{S}^\prime(\mathcal{D})\cup\{\piv\}$. The design of $f^\prime$ enforces nonzero mapping probability only when $\widetilde\piv\in\mathcal{S}^\prime(\mathcal{D})$ or $\widetilde\piv=\piv(\mathcal{D})$, while $f$ attains dataset-independent support $\mathcal{S}$.}
 Hence, the corresponding transition matrix $\Fv^\prime$ has non-zero entries only in a submatrix $\Fv^\prime \in [0,1]^{(|\mathcal{S}^\prime(\mathcal{D})|+1)\times (|\mathcal{S}^\prime(\mathcal{D})|+1)}$, whose rows and columns are indexed by $\piv(\mathcal{D})$ and the elements of $\mathcal{S}^\prime(\mathcal{D})$. This strategy reduces the number of free variables in $\Fv$ from $\binom{N-1}{K-1}^2$ to $\mathcal{O}(K^4)$.  To further simplify the optimization of this reduced matrix, we additionally impose the row-balancing constraint $\Fv^\prime{\bf 1}={\bf 1}.$ This normalization constraint requires the transition probabilities into each output state to sum to one over all reduced input states. Together with column normalization, it makes $\Fv^\prime$ doubly stochastic.

Under the above restrictions, the original constraint \eqref{eq11} becomes:
\begin{align}\label{eq16}
        \epsilon_0\geq \left|\ln\frac{f^\prime(\widetilde\piv|\piv(\mathcal{D}))}{f^\prime(\widetilde\piv|\piv^\prime))}\right|,\forall \widetilde \piv\in\mathcal{S}^\prime(\mathcal{D})\cup \{\piv(\mathcal{D})\},\forall \piv^\prime\in\mathcal{S}^\prime(\mathcal{D}).
\end{align}
Fixing an indexing order for $\{\piv(\mathcal{D})\}\cup \mathcal{S}^\prime(\mathcal{D})$, let $\piv(\mathcal{D})$ correspond to the $j^\star$-th column of $\Fv^\prime$. Then \eqref{eq16} is equivalent to
\begin{align}\label{eq17}
e^{-\epsilon_0} F_{i,j}^\prime\leq F_{i,j^\star}^\prime\leq e^{\epsilon_0} F_{i,j}^\prime,\forall j\neq j^\star, \forall 1\leq i\leq |\mathcal{S}^\prime(\mathcal{D})|+1.
\end{align}

Next, we simplify the constraints involving $\Gammav_k$ by switching to its inverse $\Gammav_k^{-1}$. Using the Schur complement on $\Gammav_k^{-1}$, we have
\begin{align}
   & \Gammav_k\succeq{\bf 0}\Leftrightarrow\Gammav_k^{-1}\succeq{\bf 0},\\
&\eqref{eq10}\Leftrightarrow\begin{bmatrix}
	\epsilon_k^2/(2\log(4/\delta)) & (\muv_k(\mathcal{D})-\muv_k(\mathcal{D}^\prime_{n,k^\prime}))^T\\
	\muv_k(\mathcal{D})-\muv_k(\mathcal{D}^\prime_{n,k^\prime}) & \Gammav_k^{-1}
\end{bmatrix}\succeq{\bf 0}.\label{eq19}
\end{align}
Therefore, we can solve for $\Gammav_k^{-1}$ directly, ensuring all relevant constraints are convex in $\Gammav_k^{-1}$.

Gathering all the above modifications, we obtain a simplified optimization problem:
\begin{subequations}
\begin{align}
    (\text{P2}):&\min_{\substack{\{\gamma_k,\Gammav_k^{-1},\epsilon_k\}_{k=1}^K\\\Fv^\prime,\epsilon_0}}  \sum_{\widetilde\piv} f(\widetilde\piv|\piv(\mathcal{D}))g(\widetilde\piv,\{\gamma_k,\Gammav_k^{-1}\}_{\forall k}) \label{eq20a}\\
    &~~\text{s.t. } 2\epsilon_k+\frac{3\gamma_k}{N_k}+\epsilon_0\leq \epsilon,\forall k\\
    &~~~~~~\eqref{eq17},\eqref{eq19},\\
    &~~~~~~\Gammav_k^{-1}\succeq {\bf 0},\gamma_k>0,\epsilon_k>0,\forall k,\\
    &~~~~~~\epsilon_0\geq0,\ \Fv^\prime{\bf 1}={\bf 1},\ (\Fv^\prime)^T{\bf 1}={\bf 1},\ [\Fv^\prime]_{i,j}\geq0,
\end{align}
\end{subequations}
where $g(\widetilde\piv,\{\gamma_k,\Gammav_k^{-1}\}_{\forall k})$ is defined in \eqref{eq_obj}.

Because the feasible region of (P2) is a subset of that for (P1), the solution to (P2) is generally suboptimal but always feasible for (P1). Nonetheless, (P2) reduces the complexity substantially, down to $\mathcal{O}(K^4+NK^2)$ \emph{convex} constraints compared to the $\binom{N-1}{K-1}^2$ entries of $\Fv$ in (P1).

\subsection{Proposed Solution to (P2)}
Although Problem (P2) is now significantly simplified to a reduced number of convex constraints, the non-convex nature of the objective still makes its optimal solution intractable. In this section, we propose a sub-optimal solution that optimizes the variables in an alternating fashion. The proposed approach first initializes a feasible guess for $\epsilon_0$ and then alternately solves for  $\{\gamma_k,\epsilon_k,\Gammav_k^{-1}\}_{k=1}^K$ and $\{\Fv^\prime,\epsilon_0\}$ until convergence. The details are listed as follows.
\begin{enumerate}
	{\color{black}
		\item \textbf{Update $\{\gamma_k,\epsilon_k,\Gammav_k^{-1}\}_{k=1}^K$:} For fixed $\epsilon_0$, the component contribution to \eqref{eq_obj} is a nonnegative weighted sum of the expected Gaussian KL divergences. Each class has its own constraint $2\epsilon_k+3\gamma_k/N_k\leq\epsilon-\epsilon_0$. Thus, minimizing the expected Gaussian KL divergence separately for every class minimizes the component contribution for any feasible mapping. We therefore optimize the variables for every class $k$ in parallel.
		
		For any fixed $\epsilon_k$, the covariance-noise contribution decreases with $\gamma_k$, so the optimal solution must satisfy
		\begin{align}
			\gamma_k=\frac{N_k}{3}(\epsilon-\epsilon_0-2\epsilon_k),\qquad 0<\epsilon_k<\frac{\epsilon-\epsilon_0}{2}.\label{eq21}
		\end{align}
Given \eqref{eq21}, the feasible range is $0<\gamma_k<N_k(\epsilon-\epsilon_0)/3$. For a prescribed number of grid points $L\geq1$, we search over
\begin{align}
\gamma_k^{(\ell)}=\frac{\ell}{L+1}\frac{N_k(\epsilon-\epsilon_0)}{3},\qquad \ell=1,\ldots,L. \label{eq21a}
\end{align}
This grid excludes both endpoints to ensure $\gamma_k>0$ and leave a positive privacy budget for the mean release. For each candidate $\gamma_k$, we optimize $\epsilon_k$ and $\Gammav_k^{-1}$ by solving the following convex subproblem for class $k$:
\begin{align}
    &\min_{\Gammav_k^{-1},\epsilon_k^2}   ~~\tr(\Sigmav_k^{-1}\Gammav_k^{-1})\nonumber\\
    &\text{s.t. } \Gammav_k^{-1}\succeq {\bf 0},0<\epsilon_k^2\leq (\frac{\epsilon-\epsilon_0}{2}-\frac{3\gamma_k}{2N_k})^2,\nonumber\\
&\begin{bmatrix}
	\epsilon_k^2/(2\log(4/\delta)) & (\muv_k(\mathcal{D})-\muv_k(\mathcal{D}^\prime_{n,k^\prime}))^T\\
	\muv_k(\mathcal{D})-\muv_k(\mathcal{D}^\prime_{n,k^\prime}) & \Gammav_k^{-1}
\end{bmatrix}\succeq{\bf 0},\forall n,k^\prime.\label{eq23}
\end{align}
For each grid candidate, we evaluate the expected component KL divergence in \eqref{eq_obj} using its $\gamma_k$ and the corresponding SDP solution, and retain the candidate with the smallest combined mean- and covariance-noise contribution. This is a semi-definite program (SDP) and can be solved by off-the-shelf solvers such as CVX \cite{cvx}. The precision matrix $\Gammav_k$ can be obtained by the inverse of $\Gammav_k^{-1}$ after the algorithm converges.}
\item \textbf{Update $\Fv^\prime$ and $\epsilon_0$}: Fixing $\{\gamma_k,\epsilon_k,\Gammav_k^{-1}\}_{k=1}^K$, the optimization of $\Fv^\prime$ and $\epsilon_0$ can be recast as 
\begin{align}
	&\min_{\Fv^\prime,\epsilon_0}\ \sum_{i=1}^{|\mathcal S^\prime(\mathcal D)|+1}g_iF^\prime_{i,j^\star}\nonumber\\
	&\text{s.t. }0\leq\epsilon_0\leq\epsilon-2\max_k\left\{\epsilon_k+\frac{3\gamma_k}{2N_k}\right\},\nonumber\\
	&e^{-\epsilon_0}F^\prime_{i,j}\leq F^\prime_{i,j^\star}\leq e^{\epsilon_0}F^\prime_{i,j},\quad\forall j\neq j^\star,\quad\forall i,\nonumber\\
	&\Fv^\prime{\bf 1}={\bf 1},\quad(\Fv^\prime)^T{\bf 1}={\bf 1},\quad[\Fv^\prime]_{i,j}\geq0,\quad\forall i,j.\label{eq24}
\end{align}
Here, $g_i=g(\widetilde\piv,\{\gamma_k,\Gammav_k^{-1}\}_{\forall k})$ for $\widetilde\piv$ that corresponds to the $i$-th row of $\Fv^\prime$ and $j^\star$ is the index where $\piv(\mathcal{D})$ correspond to the $j^\star$-th column of $\Fv^\prime$. 

The solution of \eqref{eq24} is given by the following proposition.
\proposition{\label{theorem3}
	An optimal privacy budget for \eqref{eq24} is
	\begin{align}
		\epsilon_0=\epsilon-2\max_k\left\{\epsilon_k+\frac{3\gamma_k}{2N_k}\right\}.\label{eq25}
	\end{align}
	When $|\mathcal S^\prime(\mathcal D)|>0$ and $\epsilon_0\geq 0$. Temporarily reorder the output rows so that $g_1\leq\cdots\leq g_{|\mathcal S^\prime(\mathcal D)|+1}$, while keeping the input-column indices, including $j^\star$, fixed. Compute the current-input column successively as
	\begin{align}
		F^\prime_{\ell,j^\star}&=\min\Bigg\{\frac{1}{1+|\mathcal S^\prime(\mathcal D)|e^{-\epsilon_0}},\nonumber\\
		&\qquad 1-\sum_{i=1}^{\ell-1}F^\prime_{i,j^\star}-\frac{|\mathcal S^\prime(\mathcal D)|+1-\ell}{1+|\mathcal S^\prime(\mathcal D)|e^{\epsilon_0}}\Bigg\},\nonumber\\
		&\hspace{15mm}\ell=1,\ldots,|\mathcal S^\prime(\mathcal D)|+1.\label{eq26}
	\end{align}
	Set the remaining columns to
	\begin{align}
		F^\prime_{i,j}=\frac{1-F^\prime_{i,j^\star}}{|\mathcal S^\prime(\mathcal D)|},\quad\forall j\neq j^\star,\quad\forall i.\label{eq27}
	\end{align}
	Finally, restore the original output-row order. Ties among the $g_i$ values can be resolved arbitrarily. 
	
	If $\epsilon_0=0$, all entries equal $1/(|\mathcal S^\prime(\mathcal D)|+1)$. If $\mathcal S^\prime(\mathcal D)=\varnothing$, the reduced matrix has the single entry $F^\prime_{1,1}=1$.
}
\begin{IEEEproof}
	See Appendix~\ref{appc}.
\end{IEEEproof}

\end{enumerate}

\begin{algorithm}[!t]
	\caption{The proposed DP mechanism.}
	\label{alg1}
	\begin{algorithmic}[1]
		\STATE\textbf{Input:}  
		$\epsilon$, $\delta$, $N$, $d$, $\piv(\mathcal{D})$, $\{\muv_k,\Sigmav_k\}$, $\{\muv_k(\mathcal{D}^\prime_{n,k^\prime})\}$, $\mathcal{S}^\prime(\mathcal{D})$, $I$, $L$, and $\lambda$.
		\STATE Initialize  $\epsilon_0=\epsilon/3$.
		\STATE \textbf{for} $\mbox{iter}=1,2,\cdots,I$ \textbf{ do }\\
		\STATE~~\textbf{for} $k=1,2,\cdots,K$ \textbf{ in parallel do }\\
		\STATE ~~~~	\textbf{for} each $\gamma_k$ candidate in \eqref{eq21a} \textbf{ do }
        \STATE ~~~~~ Update $\Gammav_k^{-1}$ and $\epsilon_k$ by solving \eqref{eq23};
		\STATE~~~~\textbf{end for} \\
				\STATE ~~~~	Select the solution with the minimum expected component KL divergence;\\
				\STATE~~\textbf{end for} \\
		\STATE~~Update $\Fv^\prime$ and $\epsilon_0$ by Proposition \ref{theorem3};\\
		\STATE~~\textbf{if} {$|Obj(\cdot;\mbox{iter})-Obj(\cdot;\mbox{iter}-1)|\leq 10^{-3}$}, \textbf{early stop};\\
		\STATE\textbf{end for} \\
        \STATE Compute $\Gammav_k$ by the inverse of $\Gammav_k^{-1}$,$\forall k$;\\
        \STATE Draw $\{\wv_k,\Wv_k\}$ and $\widetilde \piv$ by \eqref{eq5}--\eqref{eq6} and $f(\widetilde\piv|\piv)$.
		\STATE	\textbf{Output:} {$\widetilde\piv$ and $\{\widetilde\muv_k,\widetilde\Sigmav_k\}_{\forall k}$.}
	\end{algorithmic}
\end{algorithm}

The covariance-noise expectation in \eqref{eq_obj} is evaluated by Monte Carlo averaging over $1000$ standard-Wishart draws per component. The same draws are reused across alternating iterations and rescaled by the current $\gamma_k^{-1}$. The trace and log-determinant contributions are evaluated jointly for each draw, while the mean-noise contribution is computed analytically.

The overall procedure is summarized in Algorithm~\ref{alg1}, shown at the top of this page. {\color{black}
In our empirical tests, the final KL-divergence performance was relatively robust to different feasible initial privacy-budget allocations. We therefore use the simple initialization $\epsilon_0=\epsilon/3$. The privacy budgets are then optimized together with the perturbation parameters during the alternating updates.} {In Step~11, we include an early-stopping criterion when the change in the objective value of \eqref{eq20a} between two consecutive iterations falls below a small threshold $10^{-3}$. In practice, convergence is typically achieved within $10$ iterations.} Since the objective is bounded below by zero and each iteration produces a {non-increasing} sequence of objectives, convergence of Algorithm~\ref{alg1} is guaranteed.

{\remark[Why the proposed mechanism set matters?]{The colored Gaussian, Wishart, and randomized-mapping mechanisms admit clean DP bounds that compose directly, while their joint effect yields a simplified KL divergence expression. This structure is essential for tractable updates: the mean-noise covariances \(\{\Gammav_k^{-1}\}\) and Wishart scale matrices \(\{\gamma_k^{-1}\Iv_d\}\) enter convex subproblems, and the mapping probabilities \(\Fv^\prime\) satisfy linear constraints. The alternating scheme thus provides (i) fast convergence and (ii) interpretable privacy–utility control via dynamic reallocation of \(\{\epsilon_k\}\) and \(\epsilon_0\) across parameters.}}

\remark[Complexity Analysis]{
The primary computational cost of Algorithm~\ref{alg1} arises from solving the SDP problem in \eqref{eq23} and computing $\Fv^\prime$ via \eqref{eq26} and \eqref{eq27}. According to \cite{CO}, the SDP in \eqref{eq23} involves a $d\times d$ matrix variable and $\mathcal{O}(NK^2)$ constraints. Solving the problem via an interior-point method incurs a complexity of $\mathcal{O}(N^{3.5}K^6)$ in the regime of $N\gg d$ \cite{CO}. Meanwhile, sorting the $|\mathcal S^\prime(\mathcal D)|+1=\mathcal O(K^2)$ coefficients costs $\mathcal O(K^2\log K)$. Evaluating \eqref{eq26} with a running cumulative sum costs $\mathcal O(K^2)$, and explicitly forming the remaining columns through \eqref{eq27} costs $\mathcal O(K^4)$. Thus the overall matrix-construction cost remains $\mathcal O(K^4)$. Consequently, the overall complexity of Algorithm~\ref{alg1} is on the order of $\mathcal{O}(ILN^{3.5}K^6)$, where $I$ is the maximum alternating iterations, $L$ is the grid search length in \eqref{eq21a}. Notably, this complexity grows in a polynomial order  with the dataset size $N$, due to our restriction that the probability mapping matrix $\Fv^\prime$ only takes non-trivial entries for $\piv(\mathcal{D})$ and its adjacent frequency vectors, thus avoiding the full histogram transition matrix with $\binom{N-1}{K-1}^2$ entries.}

{
\subsection{Adaptations to Other DP Definitions}\label{sec4c}
As stated in Definition \ref{def1}, we consider \emph{label-level} DP, where adjacent datasets differ by flipping exactly one label. For a dataset with $N$ points, all label-level adjacent datasets can be enumerated explicitly, which leads to the data-dependent bounds in Theorem \ref{theorem1}.

On the other hand, we show here that this label-level DP setting is technically more challenging than record-level DP, and our analysis and method in Algorithm \ref{alg1} adapt readily to record-level DP under a uniform feature-space bound. Consider DP defined with an adjacent dataset that \emph{adds}, \emph{removes}, or \emph{replaces} one data point, and assume a uniform bound on features,
\begin{align}
\norm{\xv_n}_2 \leq B,\ \forall n, \label{eqr1}
\end{align}
for some constant $B<\infty$.

We discuss the adaptation to three sub-cases for record-level DP:

\begin{itemize}


\item \textbf{Removing one data point.} Under label flips, two coordinates of $\piv$ change, giving at most $K(K-1)$ neighboring weight vectors in \eqref{eq14}. Removing one record instead changes one coordinate of the count vector $N\piv$; renormalization by $N-1$ changes multiple coordinates of $\piv$. The neighboring weights satisfy
\begin{align}
	\mathcal S^\prime(\mathcal D)=\Big\{\piv&(\mathcal D^\prime):\ \mathcal D^\prime\text{ is a remove-one neighbor of }\mathcal D,\nonumber\\
	&\left\|(N-1)\piv(\mathcal D^\prime)-N\piv(\mathcal D)\right\|_1=1\Big\}.\label{eq14b}
\end{align}
This set has at most $K$ elements, yielding fewer neighboring weight vectors than under label flips.

\item \textbf{Changing one feature value.} Here an adjacent dataset alters one entry of some $\xv_n$ while keeping labels fixed, so all adjacent datasets share the same mixture weight $\piv$. Therefore the release of $\piv$ is inherently DP and does not require randomization. We have $\widetilde{\piv}=\piv$, and we may drop the variables $\Fv$ and $\epsilon_0$ from the optimization in Problem P$1$.

With the feature bound in \eqref{eqr1}, we also obtain a similar closed-form DP bound for the Gaussian mechanism for releasing class sample means. For an adjacent dataset $\mathcal{D}^\prime$ that changes one feature value,
\begin{align}
    \norm{\muv_k(\mathcal{D})-\muv_k(\mathcal{D}^\prime)}\leq \frac{2B}{N_k},\label{eqr4}
\end{align}
where $N_k$ is the number of points in class $k$. Therefore, $\widetilde{\muv}_k$ attains $(\epsilon_k,\delta)$-DP provided (cf. \cite[Eq.~(14)]{9799519})
\begin{align}
\frac{\epsilon_k^2 N_k^2}{2\ln (2/\delta)}\geq 4B^2\norm{\Gammav_k}_2.\label{eq10a}
\end{align}
where $\norm{\Gammav_k}_2$ is the spectral norm.
This replaces the data-dependent bound in \eqref{eq10} with a (potentially looser) data-independent one. Consequently, we obtain $K$ constraints instead of the $N(K-1)$ constraints in \eqref{eq19}:
\begin{align}
	\Gammav_k^{-1}-\frac{8B^2\ln(2/\delta)}{N_k^2\epsilon_k^2}\Iv_d \succeq {\bf 0},\ \forall k. \label{eqr5}
\end{align}
Our algorithm remains applicable; only the update of $\Gammav_k^{-1}$ in \eqref{eq23} is replaced by a simpler SDP problem with fewer constraints. This variant is substantially simpler than the setting for the label-level DP since $\piv$ needs no additional protection.


\item \textbf{Adding one bounded data point.} Replacing $N-1$ by $N+1$ and remove-one by add-one adjacency in \eqref{eq14b} gives $K$ neighboring weight vectors. The mean-sensitivity constraints are evaluated for the added record.

\end{itemize}

Taken together, these cases show that the \emph{label-level} DP defined in Definition \ref{def1}  is the \emph{most technically demanding}. Once this case is handled, adapting the approach to record-level notions entails only minor, principled modifications. Accordingly, we retain label-level DP as the default setup.
}

{\textbf{Complexity and a scalable variant.} Under the feature bound in \eqref{eqr1}, a label flip changes the class-$k$ mean by at most $2B/(N_k-1)$. Hence the neighbor constraints in \eqref{eq23} can be replaced by the $K$ sufficient conditions
	\begin{align*}
		\Gammav_k^{-1}\succeq\frac{8B^2\ln(4/\delta)}{(N_k-1)^2\epsilon_k^2}\Iv_d,\quad\forall k.
	\end{align*}
	This replacement reduces the number of constraints from \(N(K-1)\) to \(K\), leading to an overall complexity \(\mathcal O(ILK^{6})\) that is independent of \(N\). While this bound is more conservative than the data-dependent one, our experiments indicate comparable qualitative privacy–utility trends.
}

\section{Experimental Results}\label{sec5}

In this section, we present the experimental results to evaluate the performance of the proposed approach.

\subsection{Results on Synthetic Data}\label{sec5a}
We begin by assessing the performance of the proposed method on a synthetic dataset drawn from a GMM. Unless otherwise specified, the simulation setup is as follows. First, we generate a ground-truth GMM with the following parameters: the class frequency vector is drawn from a Dirichlet distribution with $K = 5$ categories and concentration parameters equal to one; the mean of each component is drawn from a uniform distribution within the range $[-10, 10]^d$, with data dimension $d = 3$; and the covariance of each component is drawn from a Wishart distribution with $d+1$ degrees of freedom and a scale matrix $\Iv_d$.

Next, we use the GMM with these parameters to generate $N = 1000$ independent and identically distributed (i.i.d.) samples to form the input dataset $\mathcal{D}$. This dataset is then fitted to an empirical GMM $p(\xv, y)$ as defined in \eqref{eq01} using the approach outlined in \eqref{eq2}. We then apply the proposed DP mechanism to compute a differentially private GMM $\widetilde{p}(\xv, y)$ using Algorithm \ref{alg1}, with a predefined $(\epsilon, \delta)$-DP requirement. Unless otherwise specified, we set $\delta = 2\times10^{-5}$ and adjust $\epsilon$ to control the privacy level. The accuracy of the released model is measured by the KL divergence between the released GMM and the non-private model, i.e., $\mathrm{KL}\left( \widetilde{p}(\xv, y) \| p(\xv, y) \right)$.

We compare the performance of the proposed method against the following existing DP mechanisms:

\begin{itemize} 

\item \textbf{i.i.d. Laplace mechanism} \cite[Algorithm 2]{7590445}: This baseline method adds i.i.d. Laplace noise to the estimated parameters $\{\pi_k, \muv_k, \Sigmav_k\}_{k=1}^K$ and then projects the perturbed parameters back onto the feasible set. As demonstrated in \cite{7590445}, the noise added is determined by the $\ell_1$ sensitivity of the individual parameters. This mechanism guarantees $(\epsilon,0)$-DP, making it a stronger privacy condition and sufficient for achieving $(\epsilon,\delta)$-DP for any $\delta>0$.

\item \textbf{i.i.d. Gaussian mechanism} \cite{dwork2006calibrating}: In this baseline method, i.i.d. Gaussian noise is added to the estimated parameters $\{\pi_k, \muv_k, \Sigmav_k\}_{k=1}^K$, with variance calibrated to the $\ell_2$ sensitivity of the parameters to satisfy the $(\epsilon,\delta)$-DP constraint. The perturbed parameters are then projected onto the feasible set, which preserves DP by post-processing. Based on the norm inequality, the $\ell_2$ sensitivity is bounded by the corresponding $\ell_1$ sensitivity, which is analytically expressed in \cite{7590445}.

\item \textbf{Colored Gaussian mechanism} \cite[Eq. (12)]{9799519}: Originally designed to enhance DP for $K$-Means clustering, this method is adapted here to protect the mean vectors of the GMM components. Unlike the standard i.i.d. Gaussian mechanism, colored Gaussian noise is added to the mean vectors $\{\muv_k\}_{k=1}^K$ according to the approach in \eqref{eq5}. The covariance matrix $\Gammav_k^{-1},\forall k,$ is optimized by minimizing the MSE between the perturbed and true means, subject to the $(\epsilon, \delta)$-DP constraint.
\item\textbf{Rank-1 Singular Multivariate Gaussian (R1SMG)} \cite{ji2024less}: {This baseline calculates a Gaussian noise according to  \cite[Eq. (9)]{ji2024less} and adds the noise to the class sample means $\{\muv_k\}$. The covariance and weight mechanisms remain as in the above baselines.}
\end{itemize}

For a fair comparison, {we ensure that all the DP mechanisms above satisfy the $(\epsilon, \delta)$-DP with the same $\epsilon$ and $\delta$ under the \emph{same} accounting framework with $\delta=2\times10^{-5}$}. Note that Laplace mechanisms typically achieve $(\epsilon,0)$-DP, which is a stronger condition and, therefore, sufficient to guarantee $(\epsilon,\delta)$-DP for any $\delta>0$. All results are averaged over $100$ Monte Carlo trials unless otherwise specified. { For our approach, we set the parameters $L=10$ and $\lambda=10^{-3}$. 
}{For every experiment, we tune the sensitive hyperparameters of each baseline algorithm to ensure stable performance and a fair comparison. Whenever we find that a baseline is sensitive to a hyperparameter under different simulation settings, we tune that hyperparameter via grid search and report the best-performing configuration.}

\begin{figure}[!t]
		\centering
		\includegraphics[width=2.9 in]{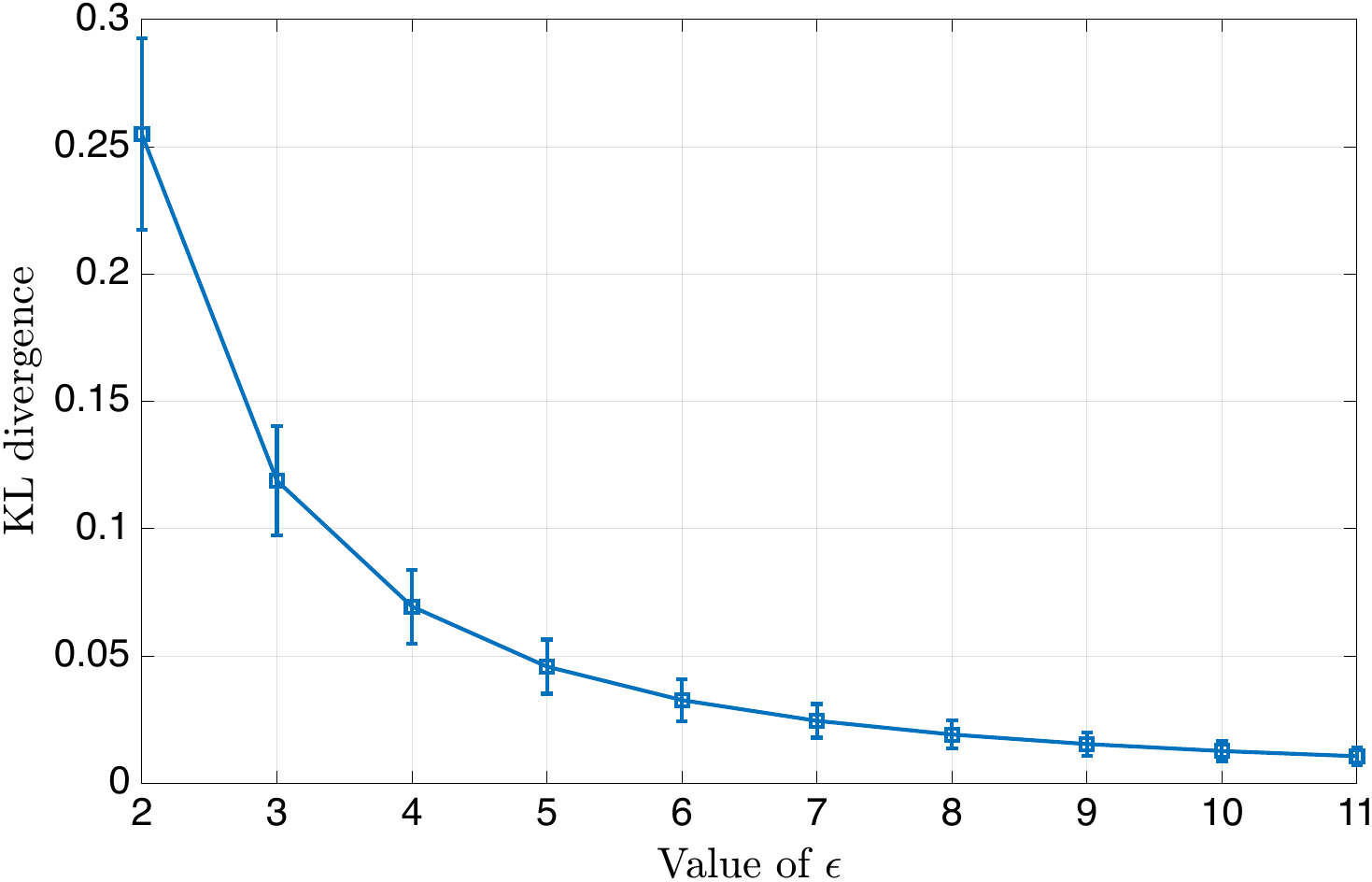}
		\caption{Average KL divergence and confidence interval for the proposed method {across $200$ Monte Carlo trials}. The privacy level is represented by the value of $\epsilon$.}
		\label{fig_CI}
\end{figure}

\begin{figure}[!t]
	\centering
	\includegraphics[width=2.9 in]{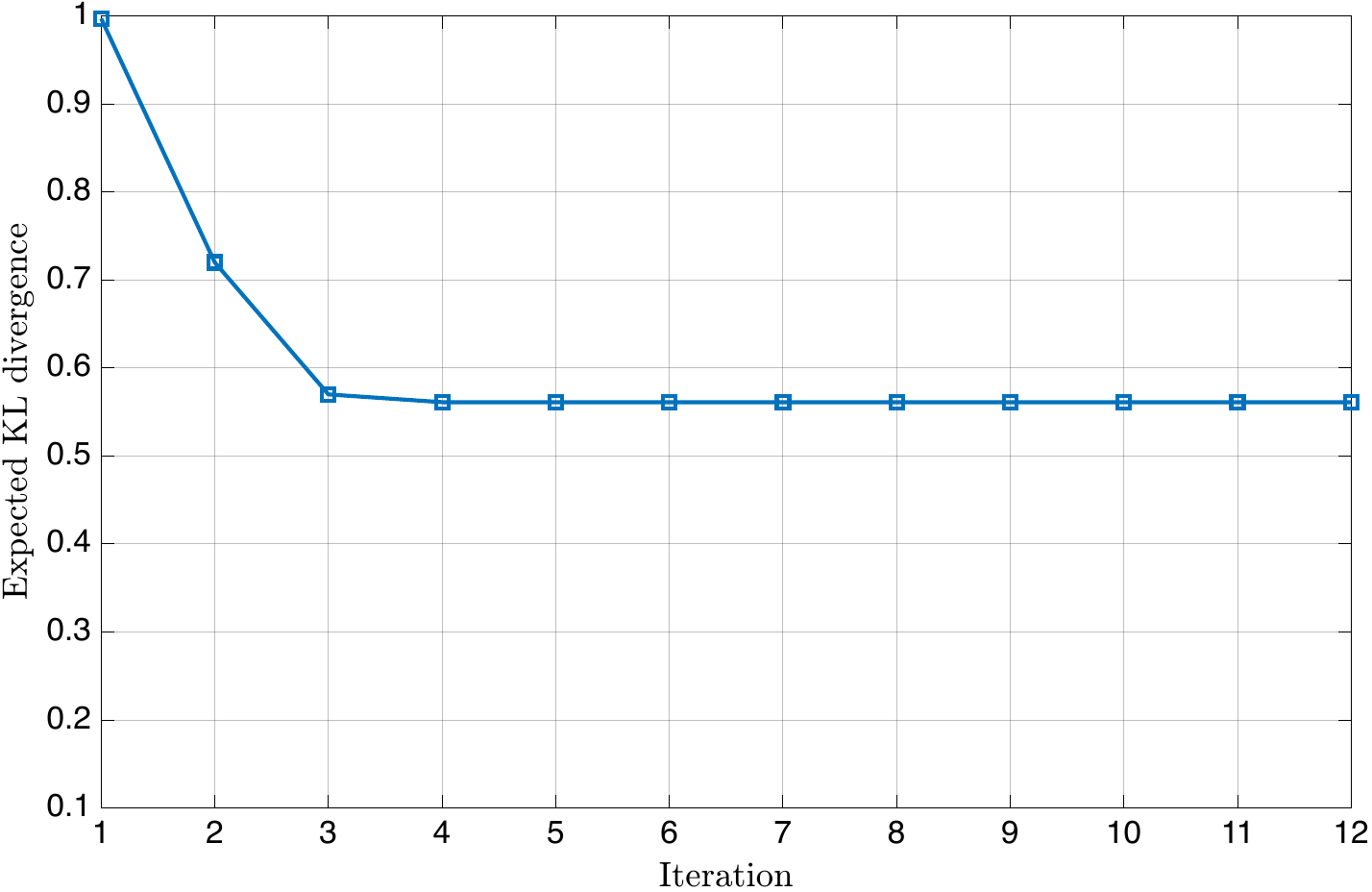}
	\caption{Expected KL divergence of the proposed algorithm versus the algorithm optimization iteration with $\epsilon=1$.}
	\label{fig_ite}
\end{figure}

\begin{figure}[!t]
	\centering
	\includegraphics[width=2.9 in]{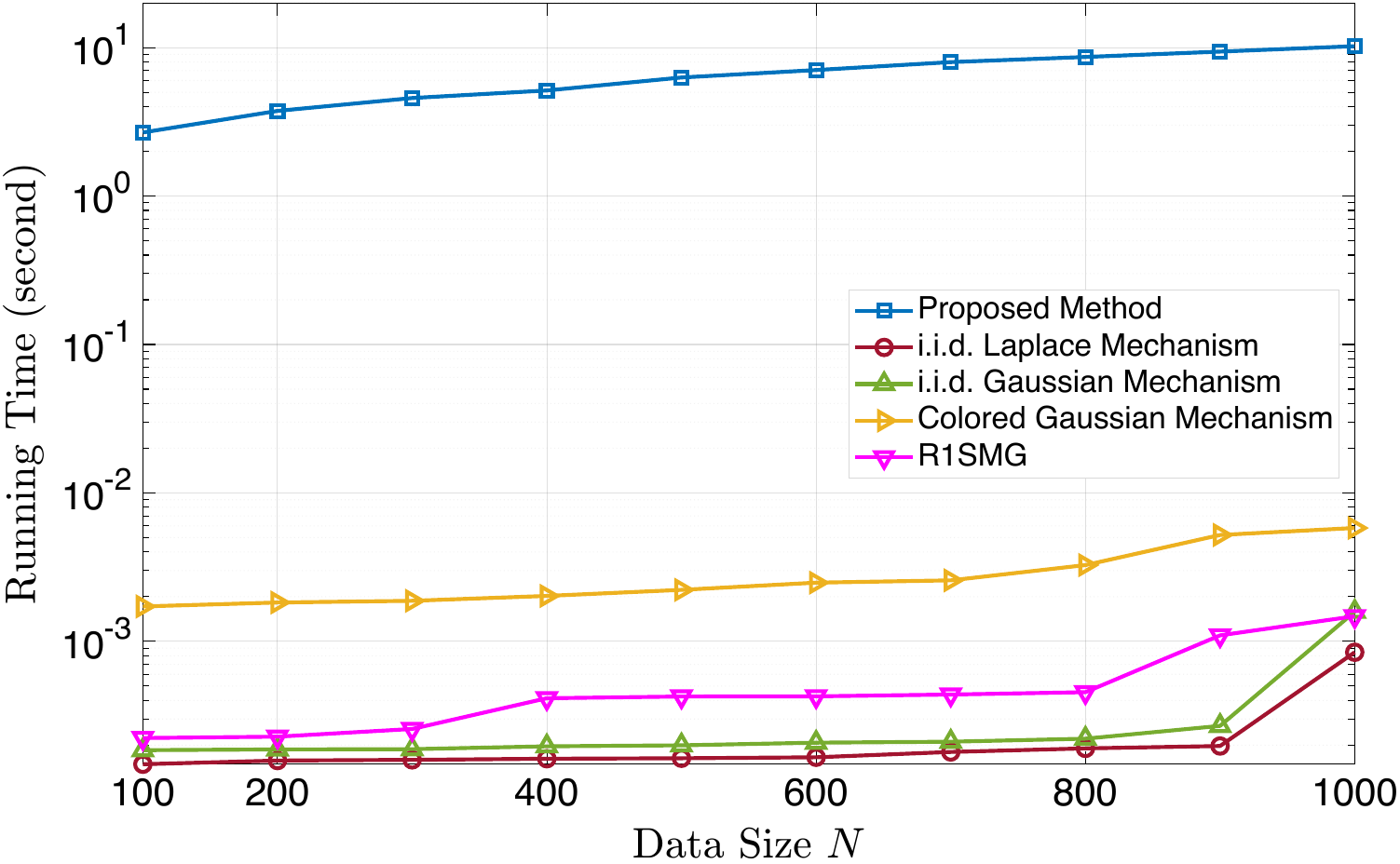}
	\caption{Running time (in seconds) of the proposed algorithm and the baselines versus the data size $N$.}
	\label{fig_runtime}
\end{figure}

\begin{figure}[!t]
		\centering
		\includegraphics[width=2.9 in]{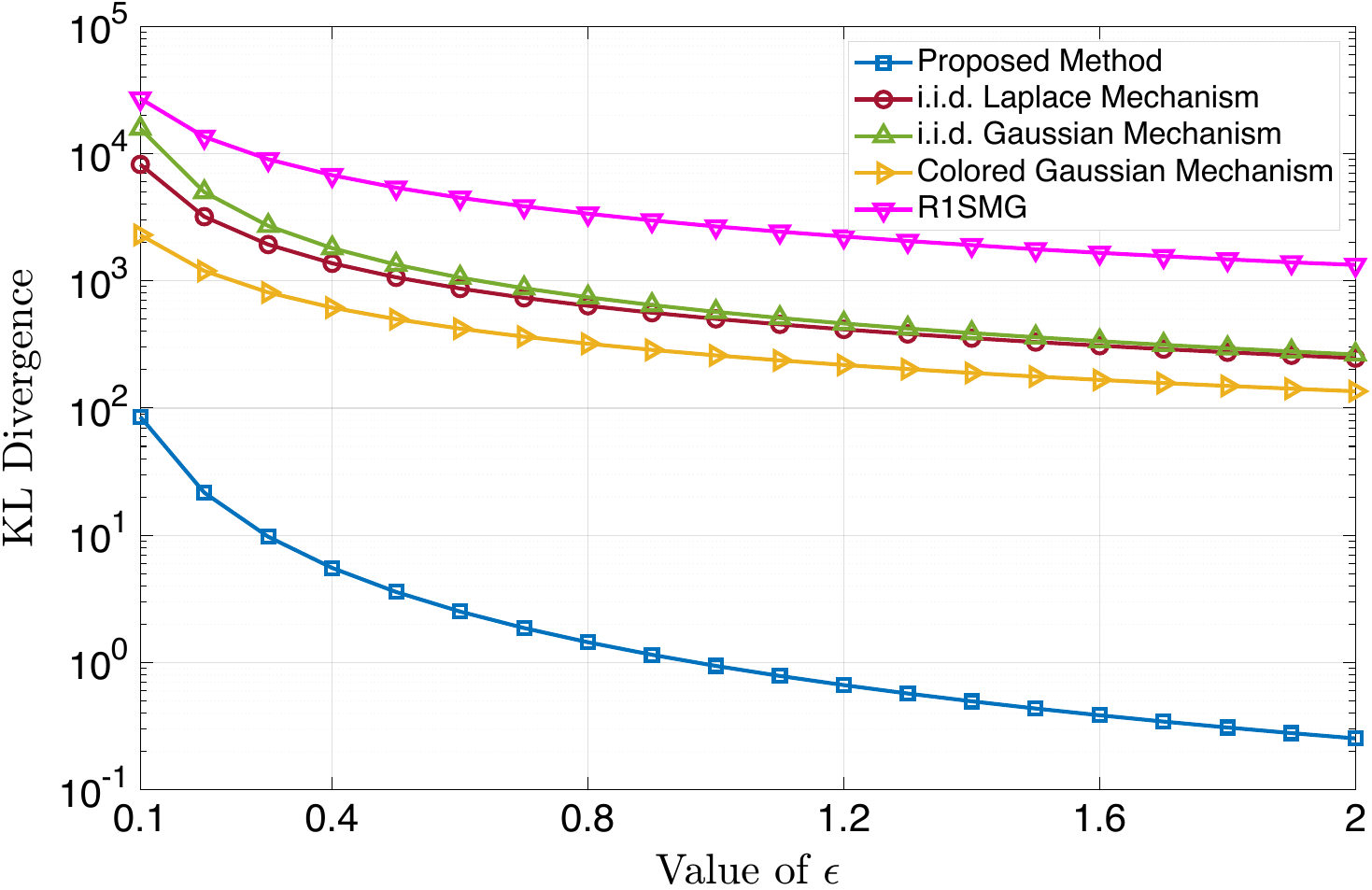}
		\caption{KL divergence versus the privacy level in terms of the value of $\epsilon$.}
		\label{fig_epsilon}
\end{figure}

We first plot the performance of the proposed method with $200$ Monte Carlo trials. Figure \ref{fig_CI} shows the mean value of the KL divergence as well as the 95\% confidence interval (error bar) under different privacy levels. The simulation is run in MATLAB R2025b on a macOS desktop (Apple M3 Ultra, 20-core CPU, 96 GB memory). We set the random seed to $42$, the smoothing parameter to $\lambda=10^{-3}$, and the early-stopping threshold to $10^{-3}$. SDP subproblems are solved in CVX \cite{cvx} using MOSEK as the default solver with default solver hyperparameters. In practice, the proposed algorithm typically terminates within $10$ iterations and each trial finishes in $2$--$10$ seconds. {\color{black}Figure \ref{fig_ite} plots the objective value (i.e., the expected KL divergence) of the proposed method versus the iteration index, with the early-stopping criterion in Algorithm \ref{alg1} disabled. The proposed method converges within five iterations. Figure \ref{fig_runtime} compares its runtime with those of the baseline methods for different dataset sizes $N$. The proposed method requires approximately $2$--$10$ seconds per run, with most of the computation time spent solving the SDP subproblems in CVX. The baseline methods admit closed-form solutions and therefore require substantially less computation time. Nevertheless, the proposed method maintains a practical runtime over the tested dataset sizes, and its runtime increases with $N$, consistent with the complexity analysis.}

We examine the utility-privacy trade-off for releasing GMMs. Figure \ref{fig_epsilon} plots the average KL divergence of the DP method under varying privacy constraints for different algorithms, as the value of $\epsilon$ changes. 
As expected, a smaller $\epsilon$ (stronger privacy) requires injecting more artificial noise, leading to greater distortion in the GMM parameters and hence a higher KL divergence. The proposed approach consistently achieves a significantly smaller KL divergence than all the other baselines, thanks to its capability to directly minimize this utility measure. { The performance of R1SMG is consistent with the calibration in \cite{ji2024less} for small ambient (or retained) dimension $d$, where the noise variance factor in the R1SMG algorithm (cf. \cite[Eq. (9)]{ji2024less}) becomes large. In the following simulations, we drop this baseline and compare our method with other mechanisms unless otherwise specified.}

\begin{figure}[!t]
		\centering
		\includegraphics[width=2.9 in]{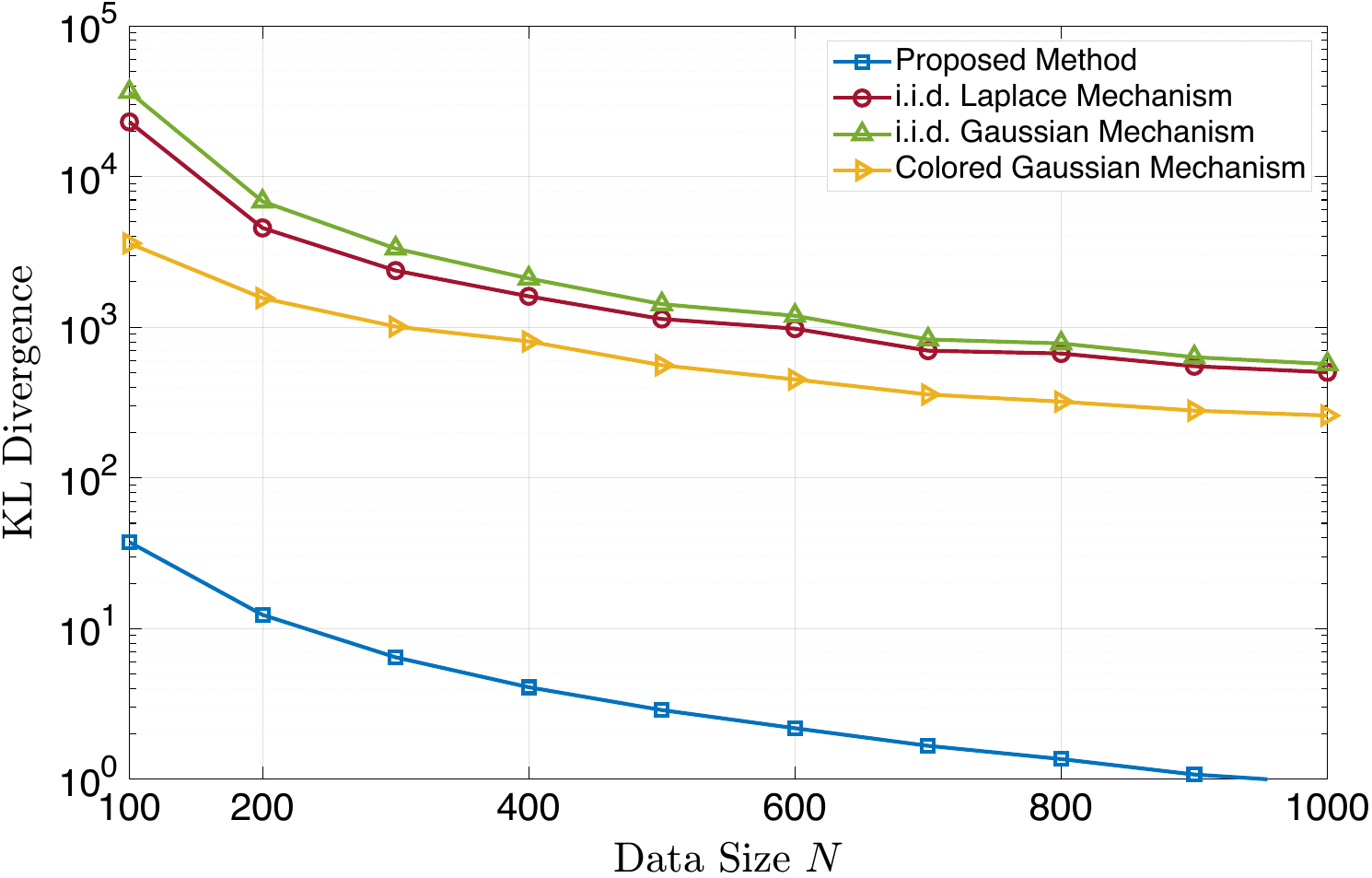}
		\caption{KL divergence versus the data size $N$ with $\epsilon=1$.}
		\label{fig_N}
\end{figure}

\begin{figure}[!t]
		\centering
			\includegraphics[width=2.9 in]{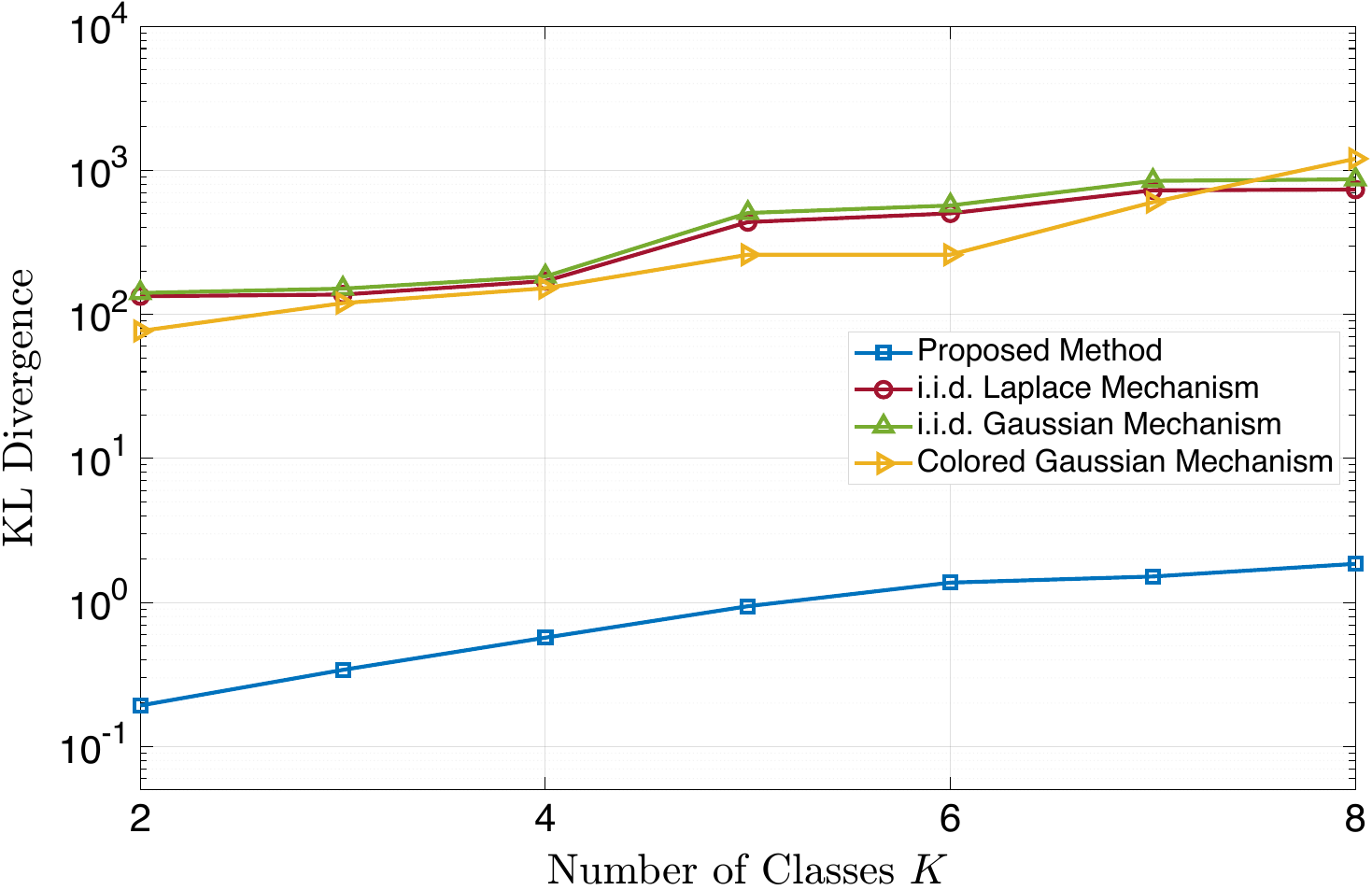}
		\caption{KL divergence versus the number of classes $K$, where the total data size is fixed to $N=1000$.}
		\label{fig_K}
\end{figure}

Next, we fix $\epsilon = 1$ and examine the KL divergence performance as a function of the total number of data points $N$, as shown in Figure \ref{fig_N}.  We see that the KL divergence decreases as $N$ increases under the same DP constraint. Intuitively, DP assesses the probability of inferring the presence or properties of a particular data point in the entire dataset. Since the data points are i.i.d. and drawn from the same GMM distribution, a larger dataset naturally has a greater capacity to obscure individual data points. As a result, it requires less noise to achieve the same level of privacy. This observation aligns with the DP analysis in Section \ref{sec3}.

Another important hyperparameter that affects the performance of differentially private GMMs is the number of data classes $K$. By fixing $N = 1000$ and $\epsilon = 1$, we plot the KL divergence performance against varying values of $K$ in Figure \ref{fig_K}. As shown in Theorem \ref{theorem1}, the per-class privacy loss for releasing the perturbed mean $\tilde{\mu}_k$ and covariance $\tilde{\Sigma}_k$ is $\epsilon_k+\epsilon_k'$, determined by the corresponding Gaussian/Wishart noise parameters. Releasing $\{(\tilde{\muv}_k,\tilde{\Sigmav}_k)\}_{k=1}^K$ across classes is accounted for by composing the releases for the source and destination classes affected by a label flip, yielding an overall cost of $2\max_k(\epsilon_k+\epsilon_k')$. {For fixed $N$, increasing $K$ reduces the average number of samples per class. Smaller classes require stronger perturbations under the same privacy budget: the Wishart privacy-loss bound scales as $3\gamma_k/(2N_k)$, and the Gaussian sensitivity bound also increases as $N_k$ decreases. In particular, under the feature bound in \eqref{eqr1}, the label-flip mean-noise bound scales as $1/(N_k-1)^2$ for fixed $\epsilon_k$ and $\delta$. Smaller classes therefore require a larger Gaussian noise covariance $\Gammav_k^{-1}$ and a larger Wishart scale $\gamma_k^{-1}$, increasing the expected KL divergence.}
This analysis is confirmed by the results in Figure \ref{fig_K}. However, the increase in KL divergence for the proposed approach is much less sensitive to $K$ than in the baseline methods, highlighting the robustness of the proposed approach.

\begin{figure}[!t]
		\centering
		\includegraphics[width=2.9 in]{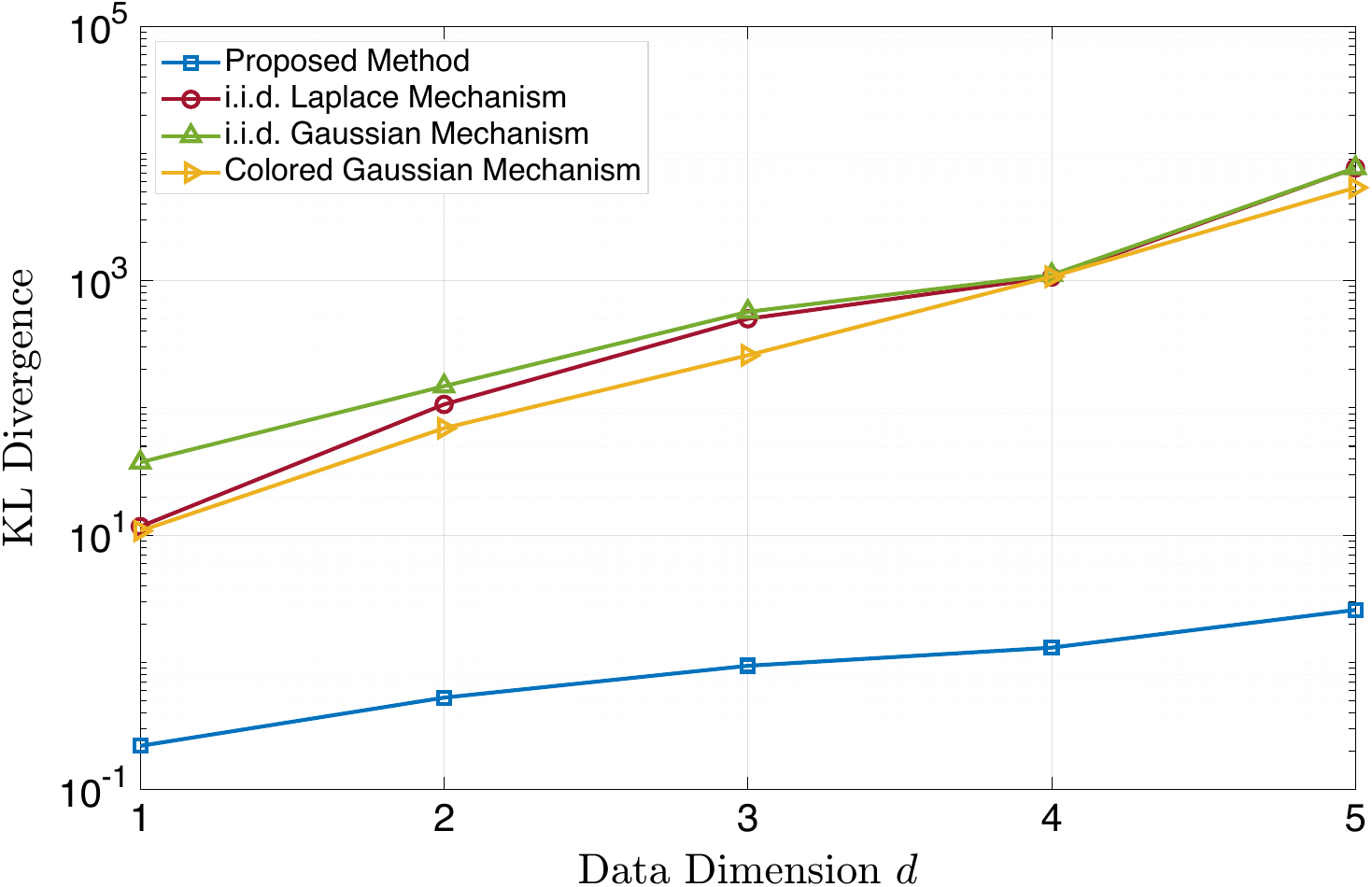}
		\caption{KL divergence versus the dimension of the data points $d$ with $K=5$ and $N=1000$.}
		\label{fig_d}
\end{figure}

\begin{figure}[!t]
        	\centering
		\includegraphics[width=2.9 in]{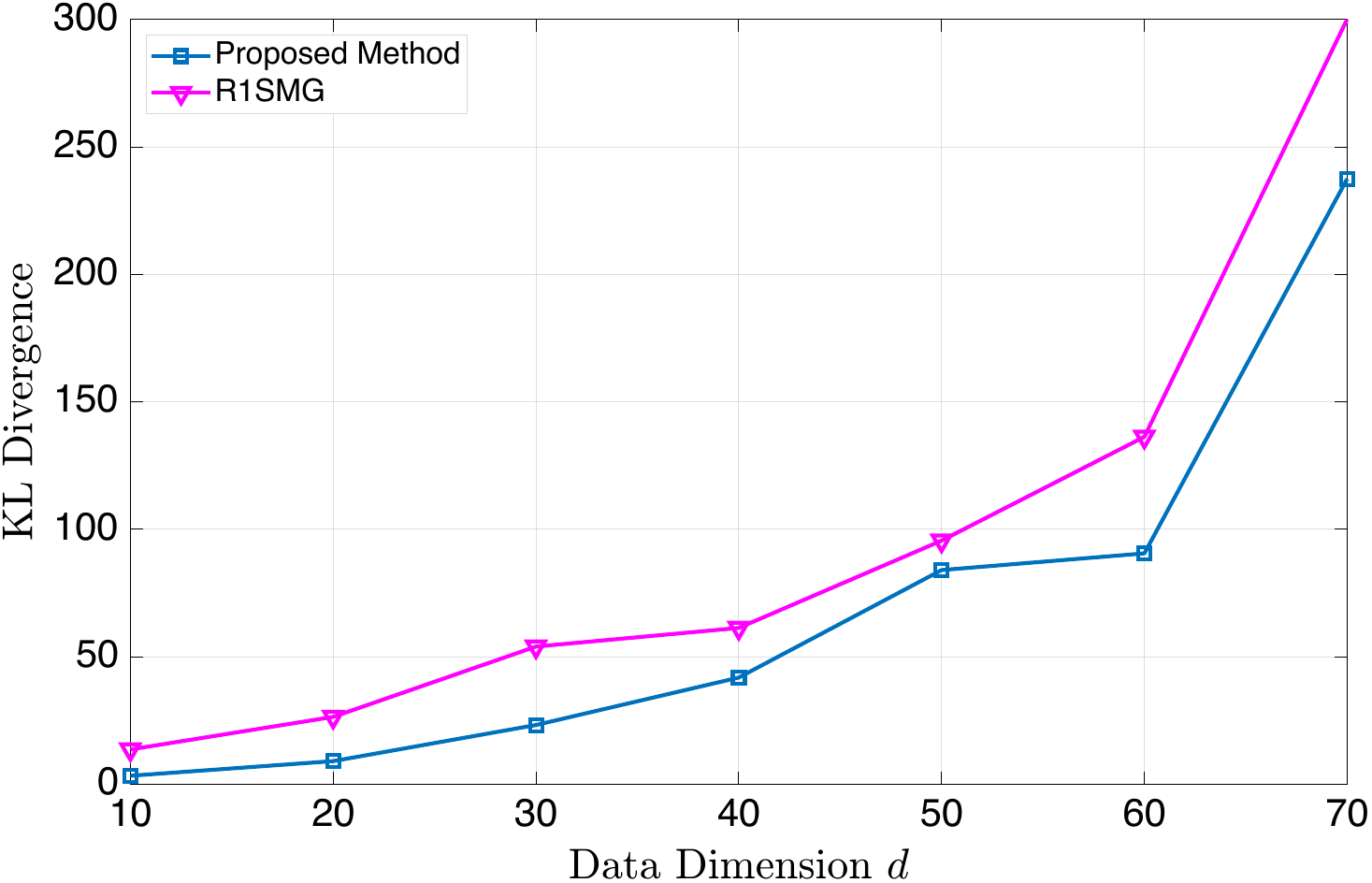}
        	\caption{KL divergence versus the data dimension $d\in [10,70]$.}
        	\label{fig_d_large}
\end{figure}

Finally, we evaluate the KL divergence with respect to the data dimension $d$ in Figure \ref{fig_d}. Moreover, in Figure \ref{fig_d_large} we vary the data dimension $d$ over $[10,70]$. From Theorem \ref{theorem1} and Proposition \ref{theorem2}, it can be seen that under the same statistics, both the KL divergence and the privacy loss increase with $d$. When $d>10$ R1SMG becomes the strongest \emph{among the baseline methods} considered in this work: it achieves a KL divergence below $3\times 10^2$, whereas the other baselines yield KL divergence values above $10^{5}$ (and are therefore omitted from the plot for readability). Importantly, the proposed method remains competitive and consistently attains a slightly smaller KL divergence than R1SMG across the entire sweep. We attribute this gain to the fact that our framework \emph{jointly} optimizes the privatization mechanisms for all GMM parameters (mixture weights, component means, and covariances) under a single expected KL-divergence objective. {Intuitively, due to the curse of dimensionality, data points become sparsely distributed in high-dimensional spaces, increasing their susceptibility to privacy risks as adjacent datasets become easier to distinguish. As a result, more perturbations are required to maintain the same level of DP. As shown in Figures \ref{fig_d} and \ref{fig_d_large}, the accuracy of all DP mechanisms significantly deteriorates as $d$ increases, demonstrating the sensitivity of differentially private GMMs to data dimensionality.}
        
\subsection{Results on the UCI Machine Learning Dataset}
\begin{figure}[!t]
		\centering
		\includegraphics[width=2.9 in]{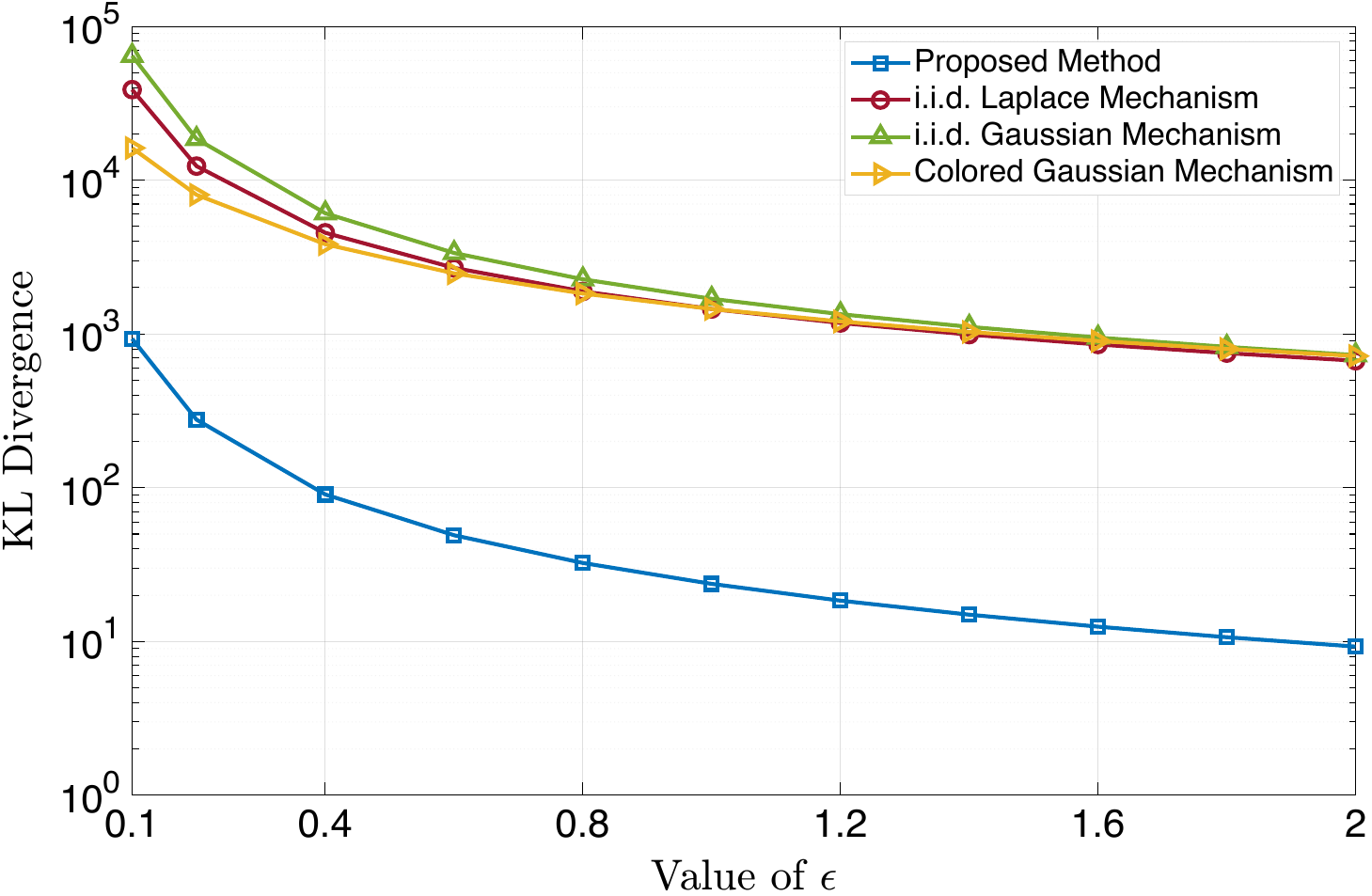}
		\caption{KL divergence versus the value of $\epsilon$ for releasing the GMM fitted to the \emph{Iris} dataset.}
		\label{fig_iris_epsilon}
\end{figure}
\begin{figure*}[!t]

	\centering
	\includegraphics[width=7 in]{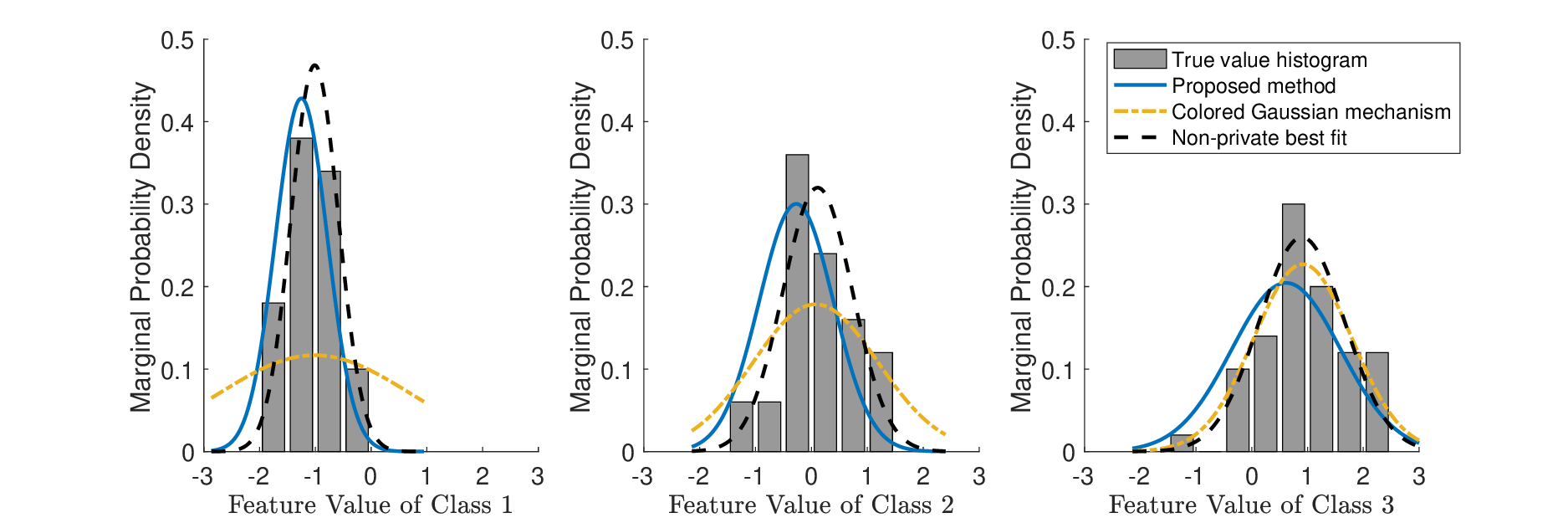}
	\caption{Marginal probability density of GMMs and empirical histograms of the true values fitted to the \emph{Iris} dataset for the first feature, i.e., the first entry of the data points $\{\xv_{n}\}_{n=1}^N$. Each subplot displays the conditional density for one class.
 } 
 \label{fig_uci}
 \hrulefill
 \end{figure*}
In this section, we evaluate the performance of the proposed DP method using real-world multi-class datasets. We employ the \emph{Iris} dataset from the UCI Machine Learning Repository \cite{iris_53}, which is a widely recognized benchmark for classification tasks. The dataset consists of $K = 3$ classes, each representing a type of iris plant, with a total of $N = 150$ instances.  Each instance is characterized by a feature vector of $d = 4$ dimensions, measuring the sepal length, sepal width, petal length, and petal width. These features are normalized to have zero mean and unit variance. Notably, the first class (Iris-Setosa) is linearly separable from the other two, while the latter two classes (Iris-Versicolor and Iris-Virginica) exhibit overlapping distributions, making them harder to distinguish.

We fit the dataset to an empirical GMM $p(\xv, y)$, as defined in \eqref{eq01}, using the method outlined in \eqref{eq2}. Subsequently, we apply the proposed DP mechanism to compute a differentially private GMM $\widetilde{p}(\xv, y)$, with a predefined $(\epsilon, \delta)$-DP constraint, where we choose $\epsilon = 2$. Figure \ref{fig_uci} visualizes the marginal probability density for the first feature (normalized sepal length) for each class. The gray bars represent the empirical histograms of the true feature values, and the black dashed lines represent the fitted GMM density in \eqref{eq2} without applying DP. The results show that the GMM produced by our method aligns more closely with the true values compared to the baseline method.

Moreover, Figure \ref{fig_iris_epsilon} illustrates the KL divergence between the differentially private GMM and the non-private model, under varying levels of privacy determined by $\epsilon$. Our approach exhibits a superior utility-privacy trade-off, achieving a smaller KL divergence under the same DP constraints. The results in Figures \ref{fig_iris_epsilon} and \ref{fig_uci} further justify the use of KL divergence as a utility metric, since a smaller divergence indicates better preservation of the fitted non-private GMM, demonstrating the effectiveness of our method in preserving data accuracy while maintaining privacy.

\subsection{Results on Image Classification}

\begin{figure}[!t]
		\centering
		\includegraphics[width=2.9 in]{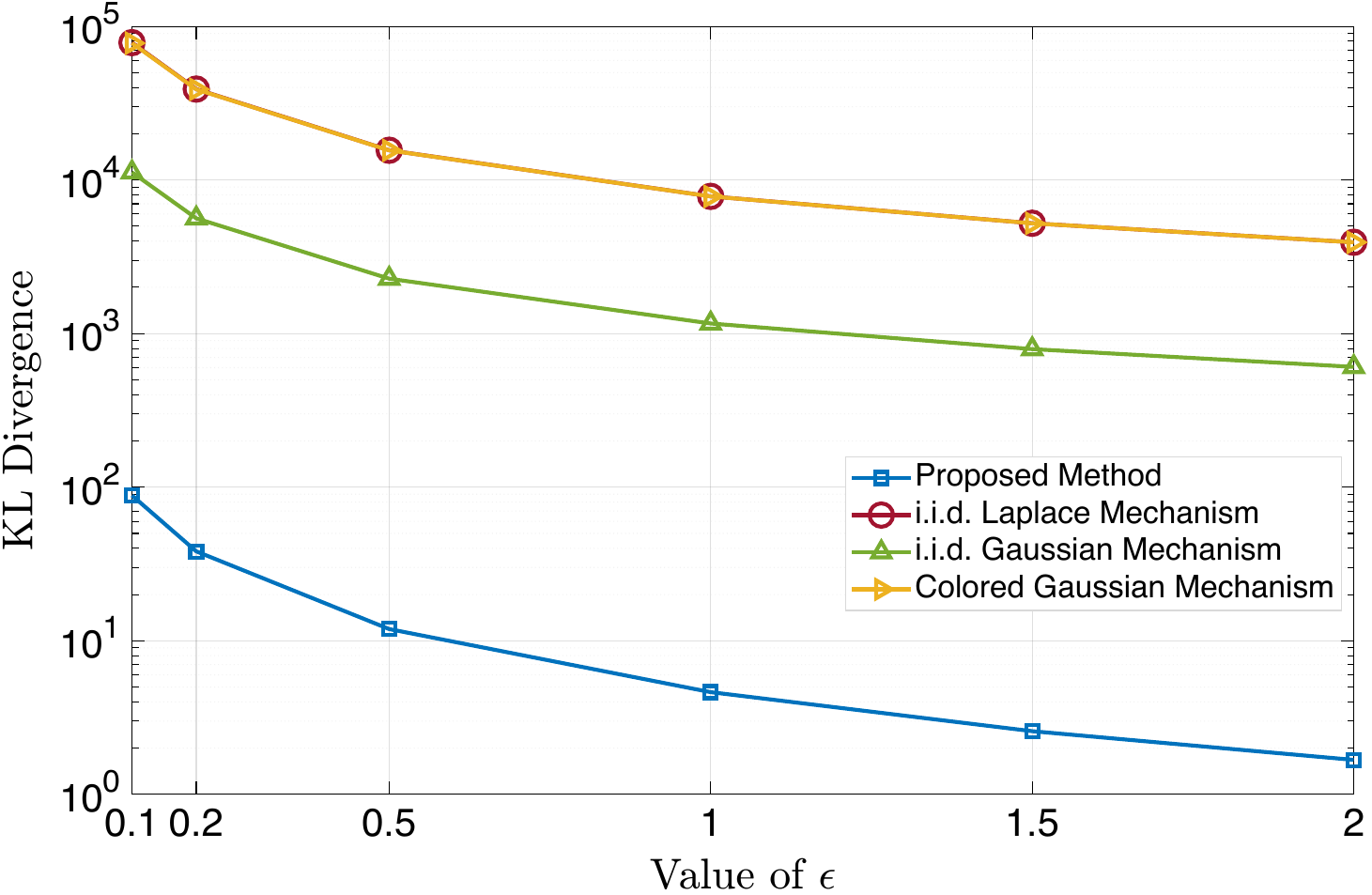}
		\caption{KL divergence versus the value of $\epsilon$ for releasing the GMM fitted to the \emph{MNIST} training data using the GMM classifier.}
		\label{fig_mnist_epsilon}
\end{figure}
\begin{figure}[!t]
	\centering
	\includegraphics[width=2.9 in]{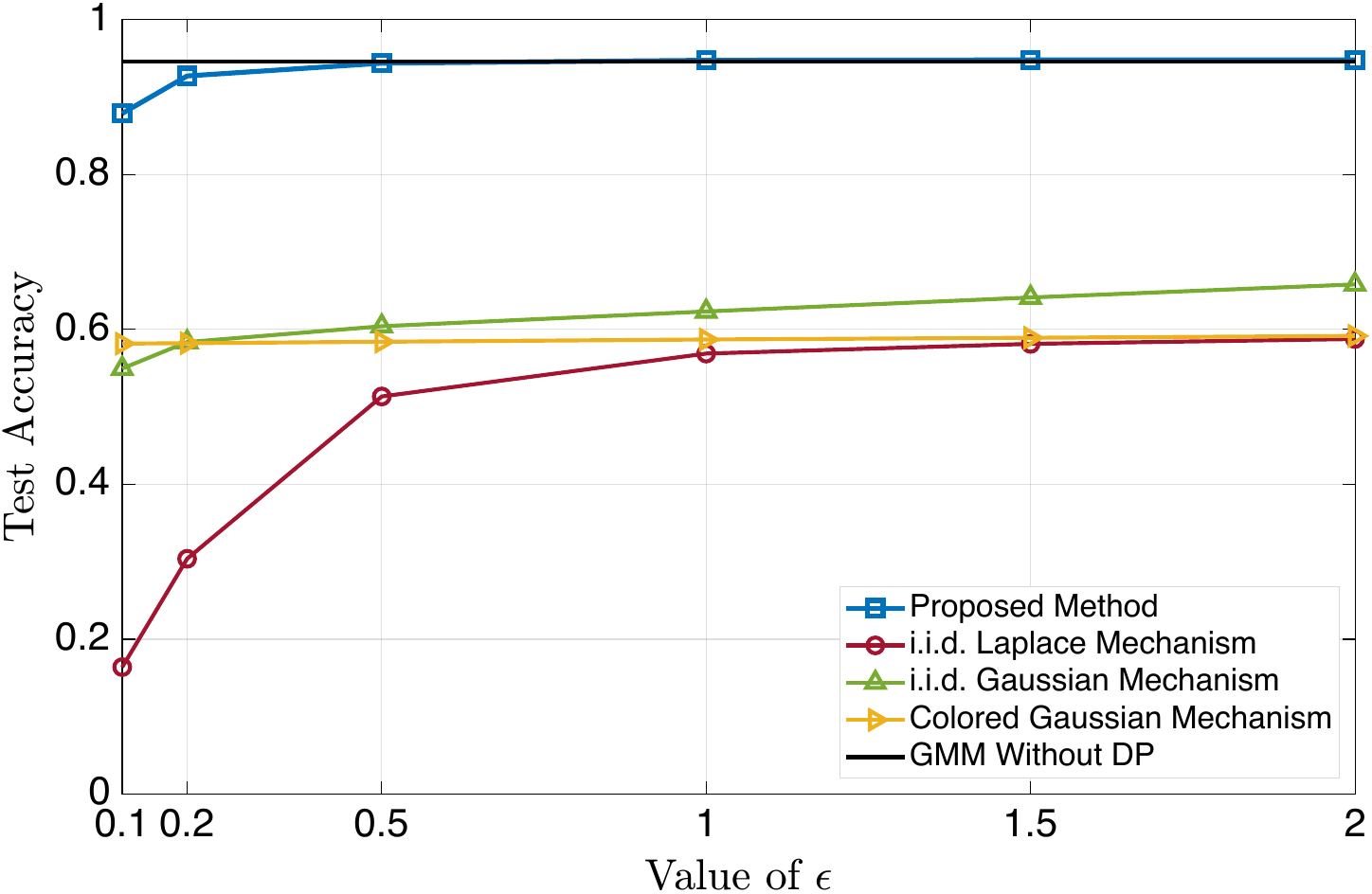}
	\caption{Test accuracy versus the value of $\epsilon$ over the \emph{MNIST} testing data using the GMM classifier.}
	\label{fig_mnist_acc}
\end{figure}
GMMs are widely adopted models for classification. In this section we explore their use in a differentially private setting for image classification. We consider the well-known \emph{MNIST} dataset \cite{deng2012mnist}, which contains $28\times28$-pixel grayscale images of handwritten digits, represented by $784$-dimensional feature vectors. We use all $N=60000$ MNIST training images and fit these data to a $10$-class GMM using the approach in \eqref{eq2} and subsequently apply the DP mechanism to enforce an $(\epsilon,\delta)$-DP constraint with $\delta=2\times10^{-5}$. We add a diagonal regularization term $10^{-3}\Iv_d$ to each fitted covariance matrix to improve its conditioning and enhance numerical stability.

Figure \ref{fig_mnist_epsilon} plots the KL divergence between the differentially private GMM and the non-private model as a function of the privacy parameter $\epsilon$. The proposed approach achieves a significantly lower divergence, indicating that the released GMM classifier closely approximates the non-private model. Furthermore, we evaluate classification performance by predicting the labels of $10^4$ test images, where each test vector is assigned to the class with the largest posterior probability under the released GMM. This maximum a posteriori (MAP) rule combines the class-conditional Gaussian density with the released mixture weight. Figure \ref{fig_mnist_acc} plots the test accuracy of different classifiers, measured by the fraction of correctly predicted labels in the range of $[0,1]$. We see that our method achieves an accuracy close to the non-private baseline, while the DP baseline approaches suffer from high distortion due to added noise. These results validate the effectiveness of using KL divergence as a utility metric, as smaller divergence indicates better preservation of the fitted non-private GMM and is accompanied by higher test accuracy in these experiments.

\begin{figure}[!t]
		\centering
		\includegraphics[width=3.2 in]{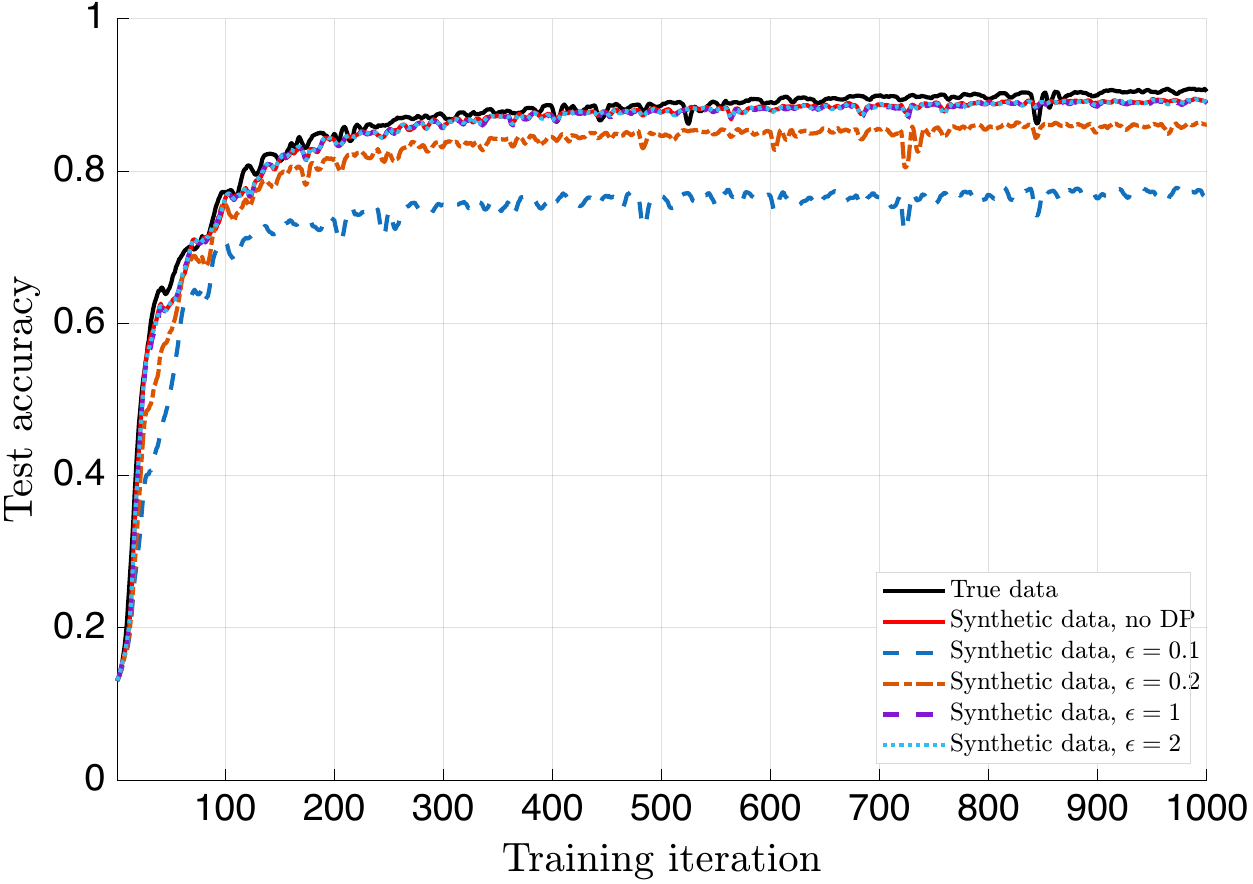}
		\caption{Test accuracy over the \emph{MNIST} testing data using the MLP model trained on synthetic data.}
		\label{fig_mnist_acc_mlp}
\end{figure}

{
The above simulation demonstrates the accuracy of DP-GMM for direct classification. In the following, we extend the simulation to an end-to-end use case where a DP-GMM is queried to synthesize training data for a downstream classifier on MNIST. Concretely, the data owner samples $N=5000$ training images from MNIST and releases the $10$-class GMM parameters $\{\widetilde\muv_k,\widetilde\Sigmav_k,\widetilde\piv\}_{k=1}^{10}$ under an $(\epsilon,\delta)$-DP budget using our method. We query the DP-GMM to generate $N=5000$ synthetic samples by drawing class labels according to the released private mixture weights and then sampling feature vectors from the corresponding privatized Gaussian components. By the post-processing theorem, the synthetic dataset remains $(\epsilon,\delta)$-DP with respect to the original training set.

For the downstream model, we train a $3$-layer multi-layer perceptron (MLP), fully connected $d\to256\to128\to10$ with ReLU activations and a softmax loss using stochastic gradient descent (SGD) with mini-batch size $128$, learning rate $0.01$, and $1000$ minibatch updates per training run. Testing is performed on the held-out ground-truth MNIST test set. Figure~\ref{fig_mnist_acc_mlp} reports test accuracy versus training iteration for synthetic datasets generated at different privacy levels. We also include two references: (i) a model trained on the non-private ground-truth training data (black curve) and (ii) a model trained on non-private GMM-synthetic data (red curve). The curves exhibit a clear privacy–utility trade-off: more stringent privacy (smaller $\epsilon$) induces larger KL divergence in the DP-GMM fit and reduces downstream accuracy. At the same time, the proximity between the black and red curves confirms that GMM-based synthetic data can train an effective classifier, and the consistent ordering across $\epsilon$ validates the suitability of KL divergence as a utility proxy for this task.
}

\subsection{Results on Electricity Load Profile Data}

We explore the application of differentially private GMMs in real-world scenarios, specifically within the context of power systems. Electricity load demand profiles, which represent the amount of electricity consumption over time, are essential for system analysis, anomaly detection, and fault diagnosis in power systems. These load profile data are typically collected by a utility provider and shared with an analyst for further analysis. It has been shown in [4] that the logarithmic values of load demand profiles fit well to GMMs. However, recent studies have highlighted significant privacy risks associated with releasing load profile data or their statistics, motivating the release of differentially private GMMs for load profile synthesis and analysis.

Here, we examine the performance of fitting and releasing GMMs on real-world load profile data under DP constraints. Following [4], we utilize a dataset from a real-world \emph{AMI} system, which includes data from $1409$ houses spread across 12 distribution circuits in California, USA, yielding $N = 1409$ samples. Each sample represents the half-day electricity consumption profile of a single house, with measurements taken hourly over $d = 12$ consecutive hours. These data points are then transformed by taking the logarithm of the power consumption. We then apply $K$-Means clustering [19] to group the data into $K = 6$ classes, representing different types of electricity consumers (e.g., residential, commercial, agricultural). Table \ref{table1} summarizes the number of data points in each class.

\begin{figure}[!t]
	\centering
	\includegraphics[width=2.7 in]{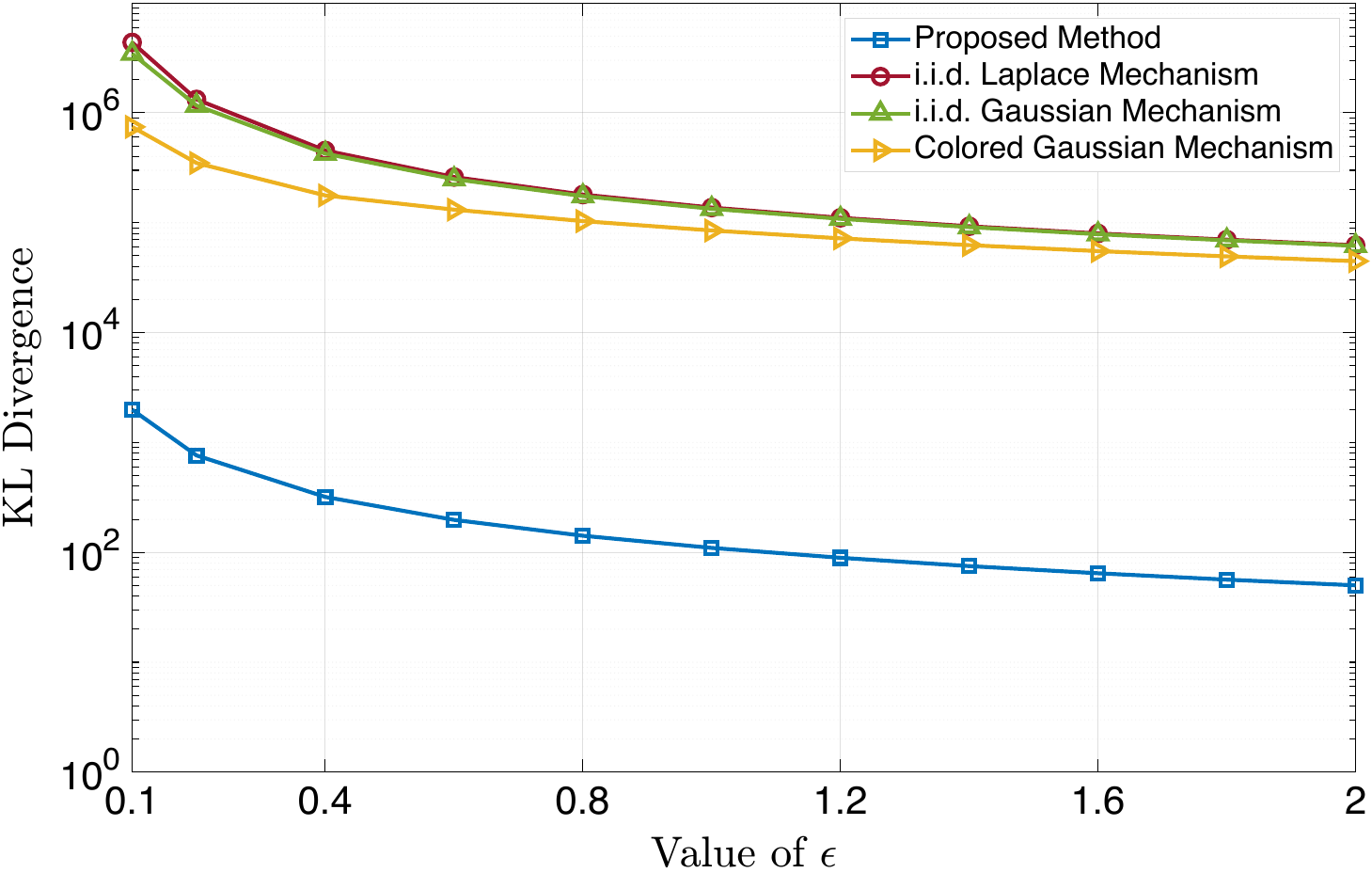}
	\caption{KL divergence versus the value of $\epsilon$ for releasing the GMM fitting to the \emph{AMI} load profile data.}
	\label{fig_AMI_epsilon}
\end{figure}

We proceed by fitting the dataset to a GMM using the method in (2) and adding artificial noise to achieve $(\epsilon, \delta)$-DP. Figure \ref{fig_AMI_epsilon} shows the KL divergence between the released GMM and the non-private model, plotted against different values of $\epsilon$, with $\delta = 10^{-5}$ fixed. The results reveal that our proposed method significantly improves the utility in terms of KL divergence compared to the baseline algorithms, effectively balancing the privacy-utility trade-off.

\begin{figure*}[!htbp]
	\centering
	\includegraphics[width=7 in,angle=90]{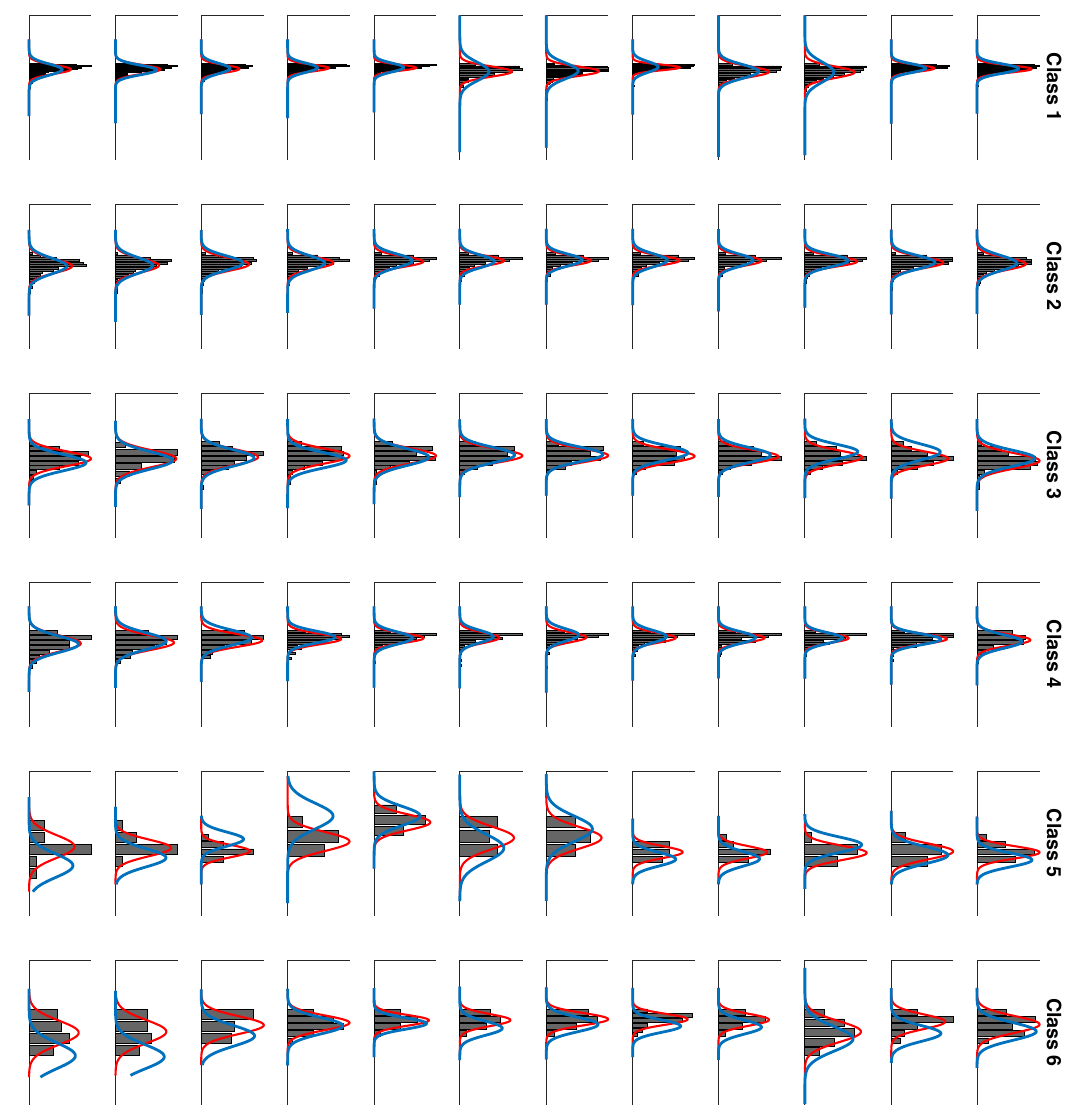}
	\caption{Entry-wise marginal probability density of GMMs fitted to the AMI load profile data for each class. The subplots in each column correspond to the marginal density of one class. In these plots, the gray bars represent the empirical histograms of the true values, the blue curves represent the marginal density of the GMM computed using the proposed DP method, and the red curves represent the marginal density of the non-private GMMs computed using (2).
	} 
	\label{fig_sub}
\end{figure*}

Next, we fix $\epsilon = 2$ and plot the marginal density of the released GMM for each dimension and class in Figure \ref{fig_sub}. For Classes 1-4, the density values closely align with the fitted GMM without DP (represented by the red curves) and the ground-truth histograms (represented by the gray bars). However, for Classes 5 and 6, the release of their parameters is more sensitive due to the limited data available, requiring significantly more noise to maintain DP. As a result, the density values from the proposed approach diverge from those of the non-private models in some dimensions for these classes. This highlights the sensitivity of differentially private GMMs to sample size, particularly when fewer data points are available.

\begin{table}[t]
	\centering
	\begin{tabular}{|c|c|}
		\hline
		Data class & \# of data points   \\ \hline
		Class $1$& $687$ \\ \hline
		Class $2$  &$428$   \\ \hline
		Class $3$ &$102$ \\ \hline
		Class $4$& $147$ \\ \hline
		Class $5$& $14$ \\ \hline
		Class $6$ & $31$\\ \hline
		Total & $1409$   \\ \hline
	\end{tabular}
	\caption{The number of data points in each class of the \emph{AMI} data.}\label{table1}
\end{table}
\subsection{Results  on Record-Level DP}
As discussed in Section IV-C in the manuscript, the proposed method extends to \emph{record-level} DP under bounded feature vectors. Here we evaluate this setting, where adjacency is defined by changing the features of a single data point. Using the synthetic dataset in Section V-A in the manuscript, we fit a GMM with an additional feature-clipping step: for each data point $\xv_n$, we apply per-record clipping $\xv_n \leftarrow \xv_n \cdot \min \{B/\norm{\xv_n}_2\}$ so that (29) holds. We then compute $(\epsilon,\delta)$-DP and adapt our algorithm exactly as described in Section IV-C in the manuscript.

\begin{figure}[!t]
	\centering
	\includegraphics[width=2.9 in]{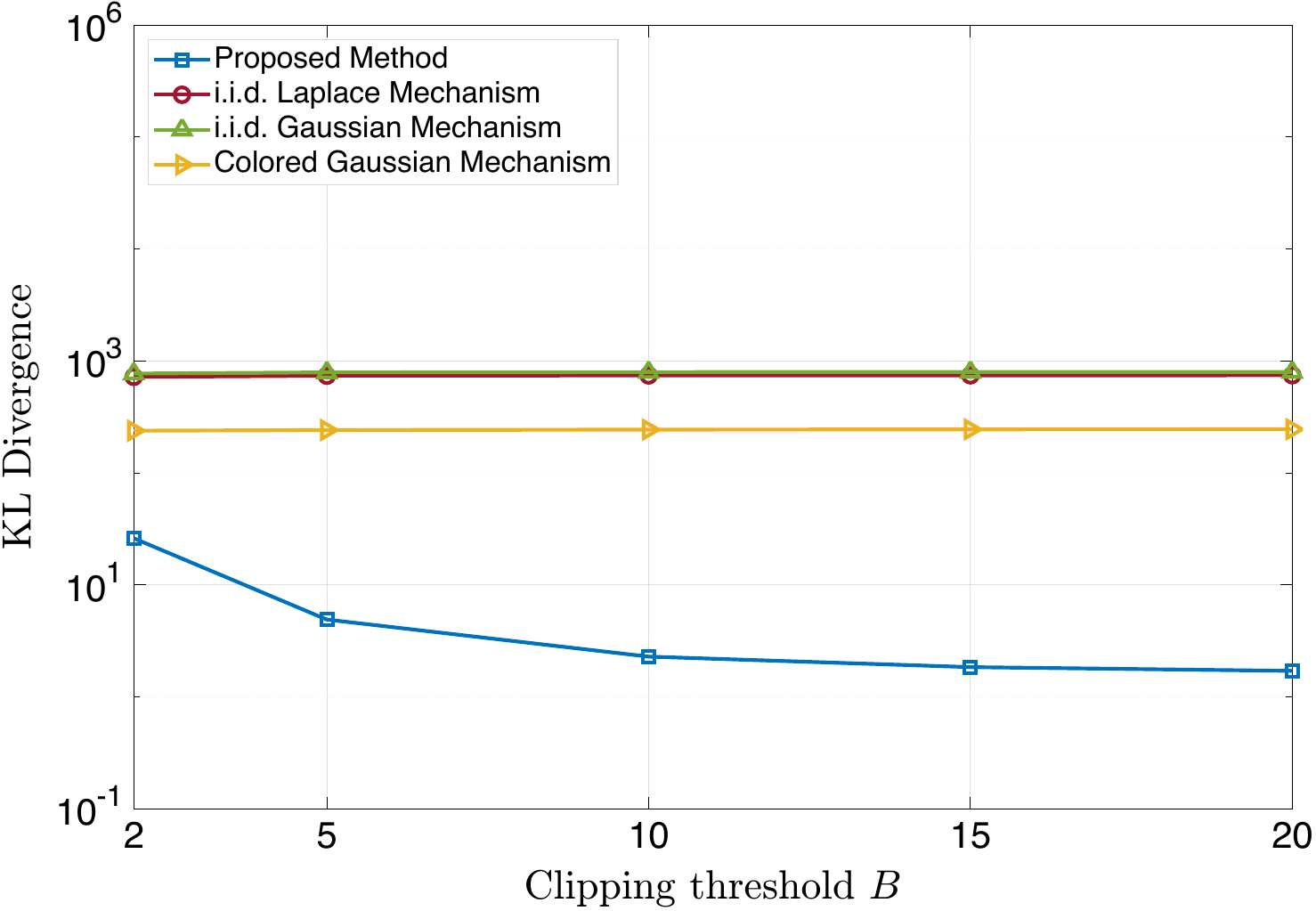}
	\caption{KL divergence versus the clipping threshold $B$ for record-level DP preservation.}
	\label{fig_B}
\end{figure}

Figure \ref{fig_B} reports the privacy--utility trade-off for varying clipping thresholds $B$ at $\epsilon=1$. For fixed class sizes and privacy budgets, the Gaussian noise-covariance lower bound in the record-level analysis of the main manuscript is proportional to $B^2$. Thus a smaller clipping bound reduces the Gaussian perturbation required by this bound.

\section{Conclusions}\label{sec6}
In this paper, we addressed the privacy protection challenge associated with releasing the parameters of GMMs, including mixture weights, component means, and covariance matrices. We proposed the use of KL divergence as a utility metric to quantify the accuracy of the released GMM, which effectively captures the combined impact of noise perturbation on individual parameters. We developed a KL-divergence-driven optimization framework for differentially private GMM release, jointly designing privacy-budget allocation and parameter perturbations. Our analysis reveals the impact of privacy budget allocation and noise statistics on DP and also offers a tractable expression for evaluating the KL divergence utility. Building on this analysis, we formulated and solved an optimization problem that minimizes the KL divergence between the released and fitted non-private models, subject to a given $(\epsilon, \delta)$-DP constraint. Experimental results on both synthetic and real-world datasets demonstrate the effectiveness of our approach, highlighting its superior performance in achieving a balance between privacy and accuracy.

\appendices

\section{Proof of Theorem 1}\label{appa}
We first characterize the DP of releasing $\{\widetilde\muv_k,\widetilde\Sigmav_k\}$ using the following established results.

\lemma[Multivariate Gaussian mechanism]{\label{lemma1} Given $\mathcal{D}$, the mechanism for publishing $\widetilde{\muv}_k$ in \eqref{eq5} satisfies $(\epsilon_k,\delta/2)$-DP given that the following inequality holds for any adjacent dataset $\mathcal{D}^\prime_{n,k^\prime}$:
	\begin{align}
		&\frac{\epsilon_k^2}{2\log (4/\delta)}\nonumber\\
		&\geq\sup_{n,k^\prime\neq k}\left\{( \muv_k(\mathcal{D})- \muv_k(\mathcal{D}^\prime_{n,k^\prime}))^T\Gammav_k( \muv_k(\mathcal{D})- \muv_k(\mathcal{D}^\prime_{n,k^\prime}))\right\}.\nonumber
	\end{align}
}
\begin{IEEEproof}
	See \cite[Theorem 3]{9799519}.
\end{IEEEproof}

\lemma[Wishart mechanism]{\label{lemma2}The mechanism for publishing $\widetilde\Sigmav_k$ in \eqref{eq6} satisfies $(\epsilon_k^\prime,0)$-DP given that
	\begin{align}
		3\gamma_k \leq 2N_k\epsilon_k^\prime.\nonumber
	\end{align}
}
\begin{IEEEproof}
	The detailed proof can be found in \cite[Theorem~4]{Jiang_Xie_Zhang_2016}. Intuitively, for a given Wishart noise realization $\overline\Wv_k$, the result is obtained via showing that 
	$$\frac{\Pr\left(\widetilde{\Sigmav}_k(\mathcal{D})=\Sigmav_k(\mathcal{D})+\overline\Wv_k\right)}{\Pr\left(\widetilde{\Sigmav}_k(\mathcal{D}^\prime)=\Sigmav_k(\mathcal{D}^\prime)+\overline\Wv_k\right)}\leq e^{\epsilon'}$$
	for any adjacent datasets $\mathcal{D}$ and $\mathcal{D}^\prime$. The left-hand side is bounded via Von Neumann’s trace inequality and the singular value inequality, leading to the constant terms in Lemma 2.
\end{IEEEproof}
{\color{black}
\lemma[Composition under a label flip]{Suppose the mean and covariance releases for class $k$ are $(\epsilon_k,\delta/2)$-DP and $(\epsilon_k',0)$-DP, respectively. A label flip affects only its source and destination classes. Hence all class-specific releases are conservatively $(2\max_k\{\epsilon_k+\epsilon_k'\},\delta)$-DP.}
\begin{IEEEproof}
	All unaffected class statistics and their fixed noise laws are identical on the two neighboring datasets. Basic composition of the source and destination mechanisms \cite{dwork2014algorithmic} gives the stated epsilon cost and $\delta/2+\delta/2=\delta$.
\end{IEEEproof}
}

Combining the above lemmas, it follows that releasing $\{\widetilde\muv_k,\widetilde\Sigmav_k\}_{k=1}^K$ adheres to $(2\max_{k=1}^K \left\{\epsilon_k+\epsilon_k'\right\},\delta)$-DP.

Next, we characterize the DP of releasing $\widetilde\piv$. From Definition 1, it can be verified that $\widetilde\piv$ satisfies $(\epsilon_0,0)$-DP if the following holds for any adjacent dataset $\mathcal{D}^\prime_{n,k^\prime}$:
\begin{align}\label{eq_appa01}
	e^{-\epsilon_0}\Pr(\widetilde\piv|\piv(\mathcal{D}^\prime_{n,k^\prime}))\leq \Pr(\widetilde\piv|\piv(\mathcal{D})) \leq e^{\epsilon_0}\Pr(\widetilde\piv|\piv(\mathcal{D}^\prime_{n,k^\prime})).
\end{align}
Combining the definition of $f(\widetilde\piv|\piv)$, it follows that the condition in \eqref{eq_appa01} is equivalent to \eqref{eq11}.

Finally, the overall DP can be obtained by using the following well-known composition theorem:
\lemma[Sequential composition]{\label{lemma4}Let $M_1$ and $M_2$ be two randomized mechanisms that apply to an input dataset $\mathcal{D}$ and satisfy $(\epsilon_1,\delta)$-DP and $(\epsilon_2,0)$-DP, respectively. Then, the sequential
	execution of them satisfies $(\epsilon_1+\epsilon_2,\delta)$-DP.
}
\begin{IEEEproof}
	See \cite{dwork2014algorithmic}.
\end{IEEEproof}
According to Lemma 4, releasing $\{\tilde\pi_k,\widetilde\muv_k,\widetilde\Sigmav_k\}_{k=1}^K$ satisfies $( 2\max_{k=1}^K\left\{\epsilon_k+\epsilon_k'\right\}+\epsilon_0,\delta)$-DP.

	\emph{Label-level adjacency and composition.}
	Under Definition~1, an adjacent dataset $\mathcal{D}^\prime$ is obtained by flipping exactly one label, i.e., moving a single sample from a source class $k_s$ to a destination class $k_t\neq k_s$ while keeping all feature vectors fixed. Therefore, only the class-$k_s$ and class-$k_t$ sufficient statistics (counts/means/covariances) can change; for any $k\notin\{k_s,k_t\}$ we have $\muv_k(\mathcal{D})=\muv_k(\mathcal{D}^\prime)$ and $\Sigmav_k(\mathcal{D})=\Sigmav_k(\mathcal{D}^\prime)$, and thus the corresponding output distributions for $(\widetilde\muv_k,\widetilde\Sigmav_k)$ are identical under $\mathcal{D}$ and $\mathcal{D}^\prime$. The Gaussian mechanism bound for releasing $\muv_k^e$ is enforced by \eqref{eq10} via Lemma 1, which uses a supremum over all admissible label flips and hence \emph{upper-bounds the worst-case (data-dependent) change} of the class-$k$ sample mean; when a flip does not involve class $k$, the mean difference is zero and the bound holds trivially. The Wishart mechanism yields the uniform per-class covariance bound for the worst-case event in Lemma 2. Since releasing $(\widetilde\muv_k,\widetilde\Sigmav_k)$ for a fixed class $k$ uses the same class-$k$ subset, sequential composition gives the per-class privacy loss $\epsilon_k+\epsilon_k^\prime$. Releasing $\{(\widetilde\muv_k,\widetilde\Sigmav_k)\}_{k=1}^K$ across all classes then gives $2\max_k\{\epsilon_k+\epsilon_k^\prime\}$. Its composing with the mixture-weight mapping (with privacy loss $\epsilon_0$) yields \eqref{eq09}.

\section{Proof of Proposition 1}\label{appb}
\begin{figure*}
	\begin{align}
		\eqref{eq08}=&\E\left[\sum_{k=1}^K\widetilde p(y=k) \int  \widetilde p(\xv|y=k)  \ln \frac{\widetilde p(y=k)  \widetilde p(\xv|y=k)}{p(y=k)p(\xv|y=k)}d\xv\right]\nonumber\\
		=&\E_{\widetilde\piv|\piv(\mathcal{D})}\left[\sum_{k=1}^K\widetilde p(y=k)\left(\ln\frac{\widetilde p(y=k)}{p(y=k)}+\E_{\{\wv_k,\Wv_k\}_{\forall k}}\left[\int  \widetilde p(\xv|y=k)  \ln \frac{\widetilde p(\xv|y=k)}{p(\xv|y=k)}d\xv\right]\right)\right]\nonumber\\
		=&\sum_{\widetilde\piv\in\mathcal{S}} f(\widetilde\piv|\piv(\mathcal{D}))\sum_{k=1}^K \tilde\pi_k\left(\ln \frac{\tilde\pi_k}{\pi_k}+\E_{\{\wv_k,\Wv_k\}_{\forall k}}\left[\int \Norm(\widetilde\muv_k,\widetilde\Sigmav_k)\ln \frac{\Norm(\widetilde\muv_k,\widetilde\Sigmav_k)}{\Norm(\muv_k,\Sigmav_k)}d\xv\right]\right).
		\label{eq_appb01}
	\end{align}
	\hrulefill
\end{figure*}
We expand the KL divergence formula in \eqref{eq08} by substituting \eqref{eq01} and \eqref{eq07}, yielding the expression in \eqref{eq_appb01} shown on top of the next page. We note that the last term on the right-hand side of \eqref{eq_appb01} is the KL divergence of two Gaussian distributions. Using the formula of the KL divergence between two multivariate Gaussian distributions, we have
{\color{black}
\begin{align}
	&\E\!\left[\mathrm{KL}\!\left(
	\Norm(\widetilde\muv_k,\widetilde\Sigmav_k)
	\middle\|\Norm(\muv_k,\Sigmav_k)\right)\right]\nonumber\\
	&=\frac12\E\!\left[
	\wv_k^T\Sigmav_k^{-1}\wv_k
	\!+\!\tr(\Sigmav_k^{-1}\Wv_k)
	\!-\!\ln\left|\Iv_d+\Sigmav_k^{-1/2}\Wv_k\Sigmav_k^{-1/2}\right|
	\right]\nonumber\\
	&=\frac12\tr(\Sigmav_k^{-1}\Gammav_k^{-1})
	+\frac{d+1}{2\gamma_k}\tr(\Sigmav_k^{-1})\nonumber\\
	&\quad-\frac12\E_{\Wv_k}\ln\left|\Iv_d+
	\Sigmav_k^{-1/2}\Wv_k\Sigmav_k^{-1/2}\right|,
	\label{eq_appb02}
\end{align}
where $\Wv_k$ has the distribution in \eqref{eq6}. Combining \eqref{eq_appb01} and \eqref{eq_appb02} proves the result.
}
\section{Proof of Proposition 2}\label{appc}

	If $\mathcal S^\prime(\mathcal D)=\varnothing$, both normalization constraints force $\Fv^\prime=[1]$, and the objective is independent of $\epsilon_0$. Hence \eqref{eq25} is an optimal budget choice whenever \eqref{eq24} is feasible. In the remainder, suppose $|\mathcal S^\prime(\mathcal D)|>0$.
	
	For a fixed feasible $\epsilon_0$, the likelihood-ratio constraints in \eqref{eq24} imply
	\begin{align*}
		e^{-\epsilon_0}F^\prime_{i,j^\star}\leq F^\prime_{i,j}\leq e^{\epsilon_0}F^\prime_{i,j^\star},\quad j\neq j^\star.
	\end{align*}
	Summing these bounds over the $|\mathcal S^\prime(\mathcal D)|$ remaining columns and using the row sum gives
	\begin{align*}
		|\mathcal S^\prime(\mathcal D)|e^{-\epsilon_0}F^\prime_{i,j^\star}&\leq1-F^\prime_{i,j^\star}\leq|\mathcal S^\prime(\mathcal D)|e^{\epsilon_0}F^\prime_{i,j^\star}.
	\end{align*}
	Consequently, every feasible current-input column satisfies
	\begin{align*}
		\frac{1}{1+|\mathcal S^\prime(\mathcal D)|e^{\epsilon_0}}&\leq F^\prime_{i,j^\star}\leq\frac{1}{1+|\mathcal S^\prime(\mathcal D)|e^{-\epsilon_0}},\quad\forall i,
	\end{align*}
	and $\sum_{i=1}^{|\mathcal S^\prime(\mathcal D)|+1}F^\prime_{i,j^\star}=1.$
	Conversely, take any current-input column satisfying these bounds and the sum constraint, and construct every remaining column using \eqref{eq27}. Each row sums to one. Every other column also sums to one because
	\begin{align*}
		\sum_{i=1}^{|\mathcal S^\prime(\mathcal D)|+1}F^\prime_{i,j}&=\frac{|\mathcal S^\prime(\mathcal D)|+1-\sum_iF^\prime_{i,j^\star}}{|\mathcal S^\prime(\mathcal D)|}=1.
	\end{align*}
	The endpoint bounds imply $e^{-\epsilon_0}F^\prime_{i,j^\star}\leq(1-F^\prime_{i,j^\star})/|\mathcal S^\prime(\mathcal D)|\leq e^{\epsilon_0}F^\prime_{i,j^\star}$, so all likelihood-ratio constraints hold as well. Thus optimizing the full reduced matrix is equivalent to minimizing $\sum_i g_iF^\prime_{i,j^\star}$ over the displayed bounded simplex.
	
	Order the rows by nondecreasing $g_i$. Initially assign every current-column entry its lower bound. Allocate the remaining probability mass in this order, filling each entry to its upper bound before proceeding to the next. At row $\ell$, the amount available is the unassigned mass after reserving the lower bound for each of the $|\mathcal S^\prime(\mathcal D)|+1-\ell$ remaining rows. This gives exactly \eqref{eq26}. The uniform vector is feasible, so the total lower-bound mass is at most one and the total upper-bound capacity is at least one; the construction therefore produces a feasible normalized column.
	
	To establish optimality, suppose $g_a<g_b$, $F^\prime_{a,j^\star}$ is below its upper bound, and $F^\prime_{b,j^\star}$ is above its lower bound. Transferring a sufficiently small mass $t>0$ from row $b$ to row $a$ preserves all column constraints and changes the objective by $t(g_a-g_b)<0$. Hence an optimal solution fills lower-cost rows before assigning residual mass to higher-cost rows. Transfers between equal-cost rows leave the objective unchanged. This proves the optimality of \eqref{eq26}; \eqref{eq27} supplies a feasible matrix completion.
	
	Finally, increasing $\epsilon_0$ decreases the lower endpoint and increases the upper endpoint of every current-column interval. The feasible set therefore expands, and the optimal objective is non-increasing in $\epsilon_0$. Choosing the largest feasible budget gives \eqref{eq25}. At $\epsilon_0=0$, both endpoints equal $1/(|\mathcal S^\prime(\mathcal D)|+1)$, yielding the uniform matrix.

\bibliographystyle{IEEEtran}
\bibliography{IEEEabrv,ref}
\end{document}

%% file: Chead.tex
\usepackage{blindtext}
\usepackage{graphicx}
\usepackage{graphicx}
\usepackage{array}
\usepackage{url}
\usepackage{algorithmic}
\usepackage{amsmath}
\usepackage{ifpdf}
\usepackage{amssymb}
\usepackage{empheq}
\usepackage{stfloats}   
\usepackage{textcomp}
\usepackage{cite}
\usepackage{multirow}
\usepackage{diagbox}
\usepackage{subcaption}
\usepackage{mathrsfs}
\let\oldFootnote\footnote
\newcommand\nextToken\relax

\renewcommand\footnote[1]{%
    \oldFootnote{#1}\futurelet\nextToken\isFootnote}

\newcommand\isFootnote{%
    \ifx\footnote\nextToken\textsuperscript{,}\fi}

\usepackage[ruled,vlined,lined,ruled,boxed]{algorithm2e}
\usepackage[dvipsnames,usenames]{color}
\usepackage[normalem]{ulem}
\usepackage{amsthm}

\usepackage{graphicx}
\usepackage{ifpdf}
\usepackage{cite}
\usepackage{enumerate}
\usepackage{enumitem}

\usepackage{array}
\usepackage{mdwmath}
\usepackage{mdwtab}
\usepackage{eqparbox}
\usepackage{bm}

\usepackage{setspace}
\usepackage{booktabs}
\usepackage[subnum]{cases}
\usepackage{rotating}

\usepackage{listings} 
\definecolor{Red}{rgb}{1,0,0}
\definecolor{Blue}{rgb}{0,0,1}
\definecolor{Olive}{rgb}{0.41,0.55,0.13}
\definecolor{Green}{rgb}{0,1,0}
\definecolor{MGreen}{rgb}{0,0.8,0}
\definecolor{DGreen}{rgb}{0,0.55,0}
\definecolor{Yellow}{rgb}{1,1,0}
\definecolor{Cyan}{rgb}{0,1,1}
\definecolor{Magenta}{rgb}{1,0,1}
\definecolor{Orange}{rgb}{1,.5,0}
\definecolor{Violet}{rgb}{.5,0,.5}
\definecolor{Purple}{rgb}{.75,0,.25}
\definecolor{Brown}{rgb}{.75,.5,.25}
\definecolor{Grey}{rgb}{.5,.5,.5}

\newcommand{\boxhead}[5]{
   \pagestyle{myheadings}
   \thispagestyle{plain}
   \setcounter{page}{1}
   \noindent
   \begin{center}
   \framebox{
      \vbox{\vspace{2mm}
    \hbox to 6.28in { {\bf #1 \hfill} }
       \vspace{6mm}
       \hbox to 6.28in { {\Large \hfill \bf #2  \hfill} }
       \vspace{6mm}
       \hbox to 6.28in { {\it #3 #4 \hfill  #5} }
      \vspace{2mm}}
   }
   \end{center}
   \markboth{#5 -- #2}{#5 -- #2}
   \vspace*{4mm}
}

\usepackage[colorlinks=true, urlcolor=black,  linkcolor=black, citecolor=black]{hyperref}
\theoremstyle{definition}
\newtheorem{theorem}{Theorem} 
\newtheorem{proposition}{Proposition}

\newtheorem{lemma}{Lemma}

\newtheorem{remark}{Remark}

\theoremstyle{definition}
\newtheorem{definition}{Definition}
\newtheorem{example}{Example}

%% file: defns.tex
\DeclarePairedDelimiterX{\infdivx}[2]{(}{)}{%
	#1\;\delimsize\|\;#2%
}

\DeclarePairedDelimiter{\norm}{\lVert}{\rVert}
\DeclarePairedDelimiter{\abs}{\lvert}{\rvert}

\def\tr{\mathop{\rm tr}\nolimits}%
\renewcommand{\Pr}{\mathscr{P}}

\newcommand{\Cv}{{\bf C}}

\newcommand{\Xv}{{\bf X}}

\newcommand{\Fv}{{\bf F}}

\newcommand{\Iv}{{\bf I}}

\newcommand{\xv}{{\bf x}}

\newcommand{\Wv}{{\bf W}}
\newcommand{\wv}{{\bf w}}

\newcommand{\Gammav}{\boldsymbol \Gamma}

\newcommand{\muv}{\boldsymbol \mu}
\newcommand{\Sigmav}{\boldsymbol \Sigma}

\newcommand{\piv}{\boldsymbol \pi}

\DeclareMathOperator\E{E}

 \def\E{\mathbb{E}}
 
 \def\Pr{\mathrm{Pr}}

\def\de \mathrm{d}

\newcommand{\Norm}{\mathcal{N}}

\def\textiid{i.i.d.\@\xspace}

\newcommand\iid{\ifmmode\text{ i.i.d. } \else \textiid \fi}

\newcommand{\beqs}{\begin{equation*}}
\newcommand{\eeqs}{\end{equation*}}
\newcommand{\beq}{\begin{equation}}
\newcommand{\eeq}{\end{equation}}